\documentclass{article}
\usepackage[margin=1in]{geometry}
\usepackage{parskip}
\usepackage{booktabs}
\usepackage{subcaption}
\usepackage[hidelinks]{hyperref}
\usepackage[ruled,vlined]{algorithm2e}
\usepackage{amsmath}
\DeclareMathOperator\arctanh{arctanh}

\usepackage{amssymb}
\usepackage{mathtools}
\usepackage{natbib}
\usepackage{subcaption}
\usepackage{graphicx}
\usepackage{color}
\usepackage{float} 
\title{On anisotropy and inhomogeneity parameter estimation using traveltimes}
\author{
Ayiaz Kaderali\footnote{%
Department of Earth Sciences, Memorial University of Newfoundland,
{\tt ayiazkaderali@gmail.com}}\,,  
Theodore Stanoev\footnote{%
Department of Earth Sciences, Memorial University of Newfoundland,
{\tt theodore.stanoev@gmail.com}}
}
\date{}
\begin{document}
\maketitle
\begin{abstract}
\noindent
We consider an anisotropic inhomogeneous model to simulate measured vertical-seismic-profile traveltimes. In this model, we assume that velocity increases linearly with depth and anisotropy is the result of elliptical velocity dependence.
Using a series of sources in a line to a single receiver, we minimize the least-squares residual between measured and modelled traveltimes to estimate the anisotropy and inhomogeneity parameters of the subsurface.
To verify the approach, we construct synthetic traveltime data for a one- and two-layer model and perform the traveltime inversion.
We justify the convergence-ensuring specifications and model-parameter restrictions that are intrinsic to the parameter estimation.
To determine the approach's practicality, we assess the reliability of results under the influence of noise.
From this assessment, we discern the noise threshold for both models and determine that the noise restriction is severely restrictive for increasing model complexity.
We estimate parameters, using the same methodology, for a real-data case and conclude with a discussion of results.
\end{abstract}
\section{Introduction}
It has been said that seismology is the most successful geophysical technique for learning about the Earth's interior~\citep[Section~9.4]{AkiRichards2002}.
A common approach used by seismologists to gain information about the stratified subsurface of the Earth is traveltime inversion.

To perform such an inversion, seismologists require at least two mathematical formalisms.
The first is  the forward problem, which, in the case of this paper, is derived using ray theory.
As such, the path of the ray that connects a source and receiver must obey Fermat's principle of stationary time.
The model must account for the material properties of the subsurface, which is satisfied by considering a velocity profile in the form of interval velocities.
The second is the quantitative modelling, which, through a numerical procedure, alters the model parameters of the forward problem to resemble a set of physical observations. 
We refer to the second formalism, as the inverse problem, {\it sensu lato}.
The choice of numerical procedure depends greatly on the structure of the forward-problem model.
In this paper, we consider an inverse problem of the least-squares type, which, in the case of traveltime inversion, seeks to minimize the difference between modelled and measured traveltimes.
The optimization technique that we consider is Newton's method in optimization, which is discussed in detail in textbooks of numerical optimization~(e.g.,~\citet[Chapter~6]{NocedalWright1999} and~\citet[Chapter~3]{Fletcher2000}).

In seismology, the combination of these mathematical formalisms forms the basis of many studies.
From~\citet{Dix1955} calculating interval velocities to~\citet{Stewart1984} using least-squares optimization techniques of traveltimes to refine interval-velocity accuracy, seismology has been used routinely to estimate layer properties of the stratified subsurface of the Earth.
Further, \citet{PujolEtAl1985} formulate multilayer ray-tracing equations for traveltime calculation of surface shot gathers.
Therein, they consider a linearized solution for the Taylor series expansion of the expressions of traveltimes and offsets, in terms of the speed and ray parameter, are truncated after the first order.
The resulting linear system is solved using the Levenberg-Marquardt algorithm, which is an iterative approach that uses the information from the gradient to determine the next iterate.
Similarly, \citet{MaoStuart1997} consider a ray-tracing multilayer model with laterally heterogeneous layers for the purpose of transmission and reflection tomography.
Therein, to minimize the misfit of their traveltime expressions, they truncate the Taylor series expansion after the first order.
Consequently, they use the Gauss-Newton algorithm, which approximates the Hessian using first-order derivatives, to minimize the traveltime misfit.

Another specification of ray theory is to model media whose velocity changes not only with depth, but also with direction, which we refer to as an anisotropic inhomogeneous medium.
Indeed, both the anisotropy and inhomogeneity of physical material have an intertwined effect on the propagation of seismic waves.
However, within the framework of mathematics, it is entirely possible to quantify their effects separately.
To this end, there is a wealth of research on the effects of anisotropy and inhomogeneity, of which we highlight~\citet{Cerveny1989,Cerveny2001},~\citet{Winterstein1990},~\citet{Helbig1994},~\citet{Slawinski2020a,Slawinski2020b}.
For the purposes of inversion, it is hopeless to attempt to invert for the model parameters that govern the symmetry class of general anisotropy.
Instead, as indicated by~\citet[p.~91]{Cerveny1989}, it is advisable to consider simpler forms of anisotropy, such as elliptically anisotropic media.

One model of this type is that of~\citet{SlawinskiEtAl2004}, wherein the defining characteristics are that the velocity of a propagating wave increases linearly with depth and exhibits an elliptical velocity dependence along the direction of propagation.
They demonstrate that the statistical significance of the anisotropy parameter, $\chi$\,, is small, but significant, and remain consistent with the weakly anisotropic assumptions discussed by~\citet{Thomsen1986}.
The model is extended by~\citet{RogisterSlawinski2005}, who use Hamilton's equations to obtain the ray-tracing equations and derive analytic solutions for the traveltime equations, which are valid for both upgoing and downgoing ray paths.
This traveltime model has been the topic of subsequent studies during the past decade.

Let us call attention to a few such studies. 
\citet{DanekSlawinski2012} use the model to perform a stochastic interpretation of geophysical data to justify the inclusion of the anisotropy parameter by means of the Bayesian Information Criterion.
Therein, to perform the traveltime inversion, they use the Metropolis-Hastings algorithm, which is a random walk algorithm that perturbs model parameters numerically to reduce the traveltime misfit.
In applying a one-layer model to synthetic and real traveltime data, they demonstrate the necessity of the anisotropy parameter to model adequately the observed traveltimes.
\citet{ZhangEtAl2017} approximate the model to calculate numerically ray paths through laterally homogeneous and heterogeneous velocity models.
By imposing triangular grids on the model subsurface, they solve the equations of motion intrinsic to the problem using the Runge-Kutta method. 
\citet{GierlachDanek2018} construct two synthetic, multilayer setups, each containing one anisotropic layer.
Using the Nelder-Mead simplex algorithm, they use the traveltime model to invert for the layer properties and, using Bayesian Information Criterion, they determine the optimal number of model parameters to model adequately the traveltimes under the influence of noise.

In this paper, we consider the mathematical model developed by~\citet{SlawinskiEtAl2004}, and extended by~\citet{RogisterSlawinski2005}, to estimate anisotropy and inhomogeneity parameters of the subsurface using vertical seismic profile traveltimes.
Its structure is as follows.
We begin by adapting the traveltime equations for multilayered models, providing an overview of Newton's method and its convergence-ensuring specifications, and establishing the requirement of logarithmic barrier functions to guarantee model-parameter restrictions. 
Then, we construct synthetic traveltime models to assess the reliability of results of the optimization under the influence of six noise profiles.
Next, we repeat the optimization for a one- and two-layer model and determine the threshold of noise that yields acceptable results for each model, which we assess using relative errors between the estimated and true model parameters.
Finally, we perform the optimization on a real-data example and conclude with a discussion of the results.
Also, this paper includes an appendix that exemplifies the control experiment of the optimization technique as well as the details of the convergence specifications and parameter restrictions.
%
\section{Formulation}
\subsection{Mathematical model}
\label{eq:abchiModel}
Let us consider a model of $N$ horizontal layers each exhibiting anisotropy and inhomogeneity.%
\footnote{
	Throughout this paper, in keeping with SI units, the units for $a$ and $b$ are m\,s$^{-1}$ and s$^{-1}$\,, respectively, for traveltime are s\,, for the ray parameter are m$^{-1}$\,s\,.
}
The inhomogeneity is a linear increase of velocity, $V(z)=a+b\,z$, where $a$ is the vertical speed of the wavefront at the top of a layer, $b$ is the measure of inhomogeneity within that layer, and $z$ is the vertical component that corresponds to depth.
The anisotropy is an elliptical velocity dependence of a wavefront~\citep{SlawinskiEtAl2004},
\begin{equation}
	\label{eq:chi}
	\chi = \frac{v_{h}^{2}-v_{v}^{2}}{2\,v_{v}^{2}}\,,
\end{equation}
where $v_h$ and $v_v$ are the horizontal and vertical velocity; for $v_h = v_v$, $\chi = 0$ and, hence, the wavefront velocity is isotropic.

We consider a series of $M$ point sources aligned in the $xz$-plane along the surface of the multilayered model and a single receiver.
Along the $j$th ray, the coordinates at the $\ell$th layer interface are given by the sequence
\begin{equation}
S_j = \{(x_{\ell,j},z_\ell)\}_{\ell=1}^{N+1}
\quad{\rm for}\quad j\in\{1\,,\dots,M\}
\,;
\end{equation}
we denote $\boldsymbol{S}$ as the $M\times1$ vector of sequences.
As such, the source coordinates are $(x_{1,j},z_1)$\,, where $x_{1,j}$ are the horizontal offsets and the vertical position $z_1 = 0$ for all $M$ sources.
The receiver coordinates are $(x_{N+1,j},z_{N+1})$\,, where $x_{N+1,j} = 0$ by design and $z_{N+1}$ is the depth of the receiver.
For the $i$th layer, $a_{i}$ is the vertical speed at the top of the layer, $b_{i}$ is the linear increase in vertical speed within the layer, and $\chi_{i}$ is the anisotropy of the layer.
Herein, $b_i>0$ corresponds to the increasing compaction of the subsurface with depth, which increases the speed of a propagating wave.
In view of physical considerations of applying the model to shale, we require $\chi_i>0$ due to the increase in the horizontal component of speed of a propagating wave, which is a directional property of shale.

Thus, the traveltime of a wave propagating along the $j$th ray is
\begin{equation}
	\label{eq:t}
	t(S_j;\boldsymbol{\beta})
	= 
	\sum\limits_{i=1}^{N}\frac{\arctanh\left(\mathfrak{p}_{i,j}\,b_i\,(x_{i+1,j}-x_{i,j})-\sqrt{1-{\mathfrak{p}_{i,j}}^{2}{a_i}^2\left(1+2\,\chi_i\right)}\right) + \arctanh\sqrt{1-{\mathfrak{p}_{i,j}}^{2}{a_i}^2\left(1+2\,\chi_i\right)}}{b_i}
	\,,
\end{equation}
where $\boldsymbol{\beta} = \left[\left\{a_i\right\}_{i=1}^{N},\left\{b_i\right\}_{i=1}^{N},\left\{\chi_i\right\}_{i=1}^{N}\right]^t$ is the $3N\times1$ vector of model parameters and
\begin{equation}
	\label{eq:p_{i,j}}
	\mathfrak{p}_{i,j}
	=
	\frac{
		2\,(x_{i+1,j}-x_{i,j})
	}{
		\sqrt{
			\left(\left(x_{i+1,j}-x_{i,j}\right)^2+\left(1+2\,\chi_i\right)\left(z_{i+1}-z_i\right)^2\right)
			\left(\left(2\,a_i+b_i\left(z_{i+1}-z_i\right)\right)^2\left(1+2\,\chi_i\right)+{b_i}^2\left(x_{i+1,j}-x_{i,j}\right)^2\right)
		}
	}
	\,,
\end{equation}
is the ray parameter in the $i$th layer, which is a conserved quantity along the ray.
Herein, we denote $\boldsymbol{t}:=\boldsymbol{t}(\boldsymbol{S};\boldsymbol{\beta})$ as the $M\times1$ vector of modelled traveltimes.
Also, let us remark that the symbol $\boldsymbol{\beta}$ should not be confused with $S$-wave speeds, which is a common use of the symbol in seismology.
To calculate the traveltime along the $j$th ray, we require a complete sequence $S_j$\,, for which the coordinates of the $N-1$ intermediate $x_{\ell,j}$ are yet unknown.
As such, for a $N$-layer model, we form the system of equations
\begin{equation}
	\label{eq:p_{i,j}system}
	\left\{
	\begin{array}{rcl}
		p_{1,j}-p_{2,j} &=& 0 \\
		p_{2,j}-p_{3,j} &=& 0 \\
		&\vdots& \\
		p_{N-1,j}-p_{N,j} &=& 0 \\
	\end{array}
	\right.
\end{equation}
and solve numerically for each $x_{\ell,j}$ for $\ell\in\{2,\dots,N\}$\,.
\subsection{Optimization technique}
\label{sec:OptTech}
To perform the optimization, we require a set of $M$ measured traveltimes that we denote by $\boldsymbol{T}$\,, which is a  $M\times1$ vector of traveltimes.
To minimize the residual between measured and modelled traveltimes, we consider the problem
\begin{equation}
	\label{eq:minf}
	\min_{\boldsymbol{\beta}} f(\boldsymbol{\beta})
	\,,
	\quad{\rm where}\quad
	f(\boldsymbol{\beta}) 
	= 
	\sum\limits_{j=1}^{M}\left(T_j - t(S_j\,;\boldsymbol{\beta})\right)^2
\end{equation}
is a twice-differentiable function, and hence, the gradient, $\boldsymbol{g}(\boldsymbol{\beta})$\,, and Hessian, $\boldsymbol{H}(\boldsymbol{\beta})$\,, of $f(\boldsymbol{\beta})$\,, exist and are continuous, whose components are
\begin{equation}
	\label{eq:g_i,H_ij}
	g(\boldsymbol{\beta})_i = \frac{\partial f(\boldsymbol{\beta})}{\partial\boldsymbol{\beta}_i}
	\quad{\rm and}\quad
	H(\boldsymbol{\beta})_{i,j} = \frac{\partial^2 f(\boldsymbol{\beta})}{\partial\boldsymbol{\beta}_i\partial\boldsymbol{\beta}_j}\,;
\end{equation} 
the former is a $3N\times1$ vector and the latter is a $3N\times3N$ matrix.
We use Newton's method in optimization, which relies on second-order Taylor series approximation of $f(\boldsymbol{\beta})$ at $\boldsymbol{\beta} = \boldsymbol{\beta}^{(k)}$\,, to form a quadratic approximation to problem~\eqref{eq:minf}, such that
\begin{equation}
	\label{eq:MultivariateVectorTaylor}
	f(\boldsymbol{\beta}^{(k)} + \alpha\,\Delta\boldsymbol{\beta}^{(k)})
	=
	\left.f(\boldsymbol{\beta})\right|_{\boldsymbol{\beta}=\boldsymbol{\beta}^{(k)}} 
	+ 
	(\alpha\,\Delta\boldsymbol{\beta}^{(k)})^t\left.\boldsymbol{g}(\boldsymbol{\beta})\right|_{\boldsymbol{\beta}=\boldsymbol{\beta}^{(k)}} 
	+ 
	\frac{1}{2}(\alpha\,\Delta\boldsymbol{\beta}^{(k)})^t\left.\boldsymbol{H}(\boldsymbol{\beta})\right|_{\boldsymbol{\beta}=\boldsymbol{\beta}^{(k)}}(\alpha\,\Delta\boldsymbol{\beta}^{(k)})
\,,
\end{equation}
where
\begin{equation}
	\label{eq:DeltaBeta^{(k)}}
	\Delta\boldsymbol{\beta}^{(k)}
	=
	-\left(\left.\boldsymbol{H}(\boldsymbol{\beta})\right|_{\boldsymbol{\beta}=\boldsymbol{\beta}^{(k)}}\right)^{-1}\left.\boldsymbol{g}(\boldsymbol{\beta})\right|_{\boldsymbol{\beta}=\boldsymbol{\beta}^{(k)}}
\end{equation}
is the Newton direction, $\alpha\in[0,1]$ is the step size, and $k$ is an iterative superscript.
To obtain the Newton direction, we set the derivative, with respect to $\Delta\boldsymbol{\beta}^{(k)}$\,, of expression~\eqref{eq:MultivariateVectorTaylor} to zero and solve for $\Delta\boldsymbol{\beta}^{(k)}$\,, where we assume $\boldsymbol{H}(\boldsymbol{\beta})$ is positive definite.
Henceforth, we adopt the notation
\begin{equation}
	\label{eq:f^k_g^k_H^k}
	f^{(k)} := \left.f(\boldsymbol{\beta})\right|_{\boldsymbol{\beta} = \boldsymbol{\beta}^{(k)}}
	\,,\quad
	\boldsymbol{g}^{(k)} := \left.\boldsymbol{g}(\boldsymbol{\beta})\right|_{\boldsymbol{\beta} = \boldsymbol{\beta}^{(k)}}
	\,,\quad
	\boldsymbol{H}^{(k)} := \left.\boldsymbol{H}(\boldsymbol{\beta})\right|_{\boldsymbol{\beta} = \boldsymbol{\beta}^{(k)}}
	\,.
\end{equation}

To determine a local solution of problem~\eqref{eq:minf}, we require an algorithm to generate a sequence of iterates~$\{\boldsymbol{\beta}\}_{k=0}^{\infty}$ that minimize the traveltime residual.
The value of~$f$ at each iterate must be such that~$f^{(k+1)} < f^{(k)}$\,, or, in other words, the sequence of iterates must lead to a strict decrease in $f$\,.
The algorithm is terminated when either a sufficient residual value is achieved or the condition of strict decrease is not met.
To generate the sequence of iterates, we consider a line search Algorithm~\ref{algo:LineSearch}, such that
\begin{equation}
\boldsymbol{\beta}^{(k+1)} = \boldsymbol{\beta}^{(k)} + \alpha^{(k)}\Delta\boldsymbol{\beta}^{(k)}
\,,
\end{equation}
which computes a step size $\alpha^{(k)}$ along the direction $\Delta\boldsymbol{\beta}^{(k)}$ to minimize $f$\,.

\begin{algorithm}[h]
	\small
	\SetAlgoLined
	For $\boldsymbol{\beta}^{(k)}$\,, calculate $f^{(k)}$\,, $\boldsymbol{g}^{(k)}$\,, $\boldsymbol{H}^{(k)}$\,, $\Delta\boldsymbol{\beta}^{(k)}$\,.
	Then, set $r=f^{(k)}$\,, $r' = r$\,, $\alpha' = 0$\,, $\Delta\alpha=0.1$\;
	\While{$r' == r$}{
		$\boldsymbol{\beta}' = \boldsymbol{\beta}^{(k)} + \alpha'\,\Delta\boldsymbol{\beta}^{(k)}$\;
		$r = \left.f(\boldsymbol{\beta})\right|_{\boldsymbol{\beta}=\boldsymbol{\beta}'}$\;
		\eIf{$r < r'$}{
			$r' = r$\;
		}{
			\eIf{$\alpha' - 2\Delta\alpha > 0$}{
				$\alpha' = \alpha' - 2\Delta\alpha$\;
			}{
				$\alpha' = 0$\;
			}
			$\Delta\alpha = \Delta\alpha/10$\;
			$\boldsymbol{\beta}' = \boldsymbol{\beta}^{(k)} + \alpha'\Delta\boldsymbol{\beta}^{(k)}$\;
			$r = \left.f(\boldsymbol{\beta})\right|_{\boldsymbol{\beta}=\boldsymbol{\beta}'}$\;
			$r'=r$\;
		}
		$\alpha' = \alpha' + \Delta\alpha$\;
		\If{$\Delta\alpha < 1\times10^{-12}$}{
			break\;
		}
		\If{$\alpha' \geq 1$}{
			break\;
		}
	}
	$\alpha^{(k)} = \alpha'$\;
	$\boldsymbol{\beta}^{(k+1)} = \boldsymbol{\beta}^{(k)} + \alpha^{(k)}\Delta\boldsymbol{\beta}^{(k)}$\;
	\caption{Backtracking line search}
	\label{algo:LineSearch}
\end{algorithm}

For Algorithm~\ref{algo:LineSearch} to produce a sequence of iterates that are globally convergent, the search direction $\Delta\boldsymbol{\beta}^{(k)}$ must be a descent direction, which is true if $\boldsymbol{H}^{(k)}$ is positive definite.
However, as indicated by~\citet[p.~136]{NocedalWright1999}, if $\boldsymbol{H}^{(k)}$ is not positive definite, or near singular, the Newton direction may lead to ascending values of $f$ or may be excessively long.
To ensure positive definiteness of $\boldsymbol{H}^{(k)}$\,, we execute the modified Cholesky factorization algorithm on $\boldsymbol{H}^{(k)}$ prior to calculating the Newton direction.
Details of this algorithm are outlined in~\citet[p.~145--149]{NocedalWright1999}, along with its use in this paper in Appendix~\ref{app:Opt_N=1_NoMods} and its necessity in Appendix~\ref{app:Barrier}. 


\subsection{Logarithmic barrier function}
\label{sec:LogBar}
The solution to problem~\eqref{eq:minf} is vector $\widehat{\boldsymbol{\beta}}$\,, which is referred to as the estimator of problem~\eqref{eq:minf}.
In unconstrained optimization, there are no restrictions on the model-parameter values that comprise $\widehat{\boldsymbol{\beta}}$\,.
However, to maintain physically meaningful results in fit-for-purpose applications of minimizing traveltime residuals, it is reasonable to impose restrictions on some, if not all, model parameters.
To do so, we introduce a logarithmic barrier function to $f$\,, which changes problem~\eqref{eq:minf} to an inequality-constrained optimization problem.
Thus, we have
\begin{equation}
	\label{eq:minf_c}
	\min_{\boldsymbol{\beta}} f(\boldsymbol{\beta})
	\quad\text{subject to}\quad
	c_i\geq0\,,\quad
	i\in\mathcal{I}\,,
\end{equation}
where $c_i$ is the set of inequality constraints and $\mathcal{I}$ is the set of integers\,.
The objective function for problem~\eqref{eq:minf_c} is
\begin{equation}
	\label{eq:P(beta)}
	P(\boldsymbol{\beta}) = f(\boldsymbol{\beta}) - \sum\limits_{i\in\mathcal{I}}\,\log l_i\,,
\end{equation}
where the summand of the latter term in $P$ is the logarithmic barrier function.
In a similar manner to notation~\eqref{eq:f^k_g^k_H^k}, we let $P^{(k)}$ denote value of the objective function at the $k$th iteration of the optimization---it is analogous to the $k$th value of $f^{(k)}$ for problem~\eqref{eq:minf}. 
Furthermore, in the context of problem~\eqref{eq:minf_c}, we abuse notation~\eqref{eq:g_i,H_ij} by stating equivalently that the components of the gradient and Hessian of $P(\boldsymbol{\beta})$ are
\begin{equation}
	g(\boldsymbol{\beta})_i = \frac{\partial P(\boldsymbol{\beta})}{\partial\boldsymbol{\beta}_i}
	\quad{\rm and}\quad
	H(\boldsymbol{\beta})_{i,j} = \frac{\partial^2 P(\boldsymbol{\beta})}{\partial\boldsymbol{\beta}_i\partial\boldsymbol{\beta}_j}
	\,.
\end{equation}

According to~\citet[Section 17.2]{NocedalWright1999}, the barrier functions have the properties such that they are infinite everywhere except in a particular domain, smooth inside the domain, and their value approaches infinity at the boundary of the domain.
To satisfy these properties, we consider the logistic function,
\begin{equation}
	\label{eq:l}
	l(x) = \frac{1}{1+\exp\left(-r\left(x-x_0\right)\right)}
	\,,
\end{equation}
where $r$ is the growth rate and $x_0$ is its center.
For large $r$\,, equation~\eqref{eq:l} resembles a step function about $x_0$ with a minimum value of zero and a maximum value of one.
Since $\log(0) = -\infty$ and $\log(1) = 0$\,, we exploit~\eqref{eq:l} to restrict the values of model parameters.
We provide further details on the use of the logistic function  as a logarithmic barrier function in Appendix~\ref{app:Barrier}.

\section{Synthetic-data testing}
Let us assess the reliability of results in using Newton's method to solve problem~\eqref{eq:P(beta)}.
For consistency, we retain the same source-receiver combinations across all executions of the algorithm.
We specify $M=139$ point sources along the surface, with offsets ranging from approximately 80\,m to 3300\,m\,, at intervals of approximately 25\,m\,, and a single receiver at a depth of approximately 1850\,m\,.
To ensure physically meaningful results, we use the logarithmic barrier function to restrict model parameters such that $b_i>0$ and $\chi_i>0$\,.

First, we set up the following control experiment.
For a known set of model parameters, $\boldsymbol{\beta}$\,, we generate sequences $\boldsymbol{S}$ and calculate traveltimes $\boldsymbol{t}(\boldsymbol{S};\boldsymbol{\beta})$\,, which we designate as the measured traveltimes $\boldsymbol{T}$\,.
Then, from a set of starting values, $\boldsymbol{\beta}^{(1)}$\,, we perform Newton's method to obtain the sequence of iterates $\{\boldsymbol{\beta}^{(k)}\}$ that minimize $f$ and, hence, solve problem~\eqref{eq:P(beta)}, whose solution is the estimator vector, $\widehat{\boldsymbol{\beta}}$\,. 
We quantify the ``goodness of fit'' by calculating the relative errors, $\delta(\widehat{\boldsymbol{\beta}}) := ((\widehat{\boldsymbol{\beta}} - \boldsymbol{\beta})/\boldsymbol{\beta})\times100\%$ 
between the observed $\widehat{\boldsymbol{\beta}}$ and expected $\boldsymbol{\beta}$\,.

\begin{figure}[h]
	\centering
	\includegraphics[width=\textwidth]{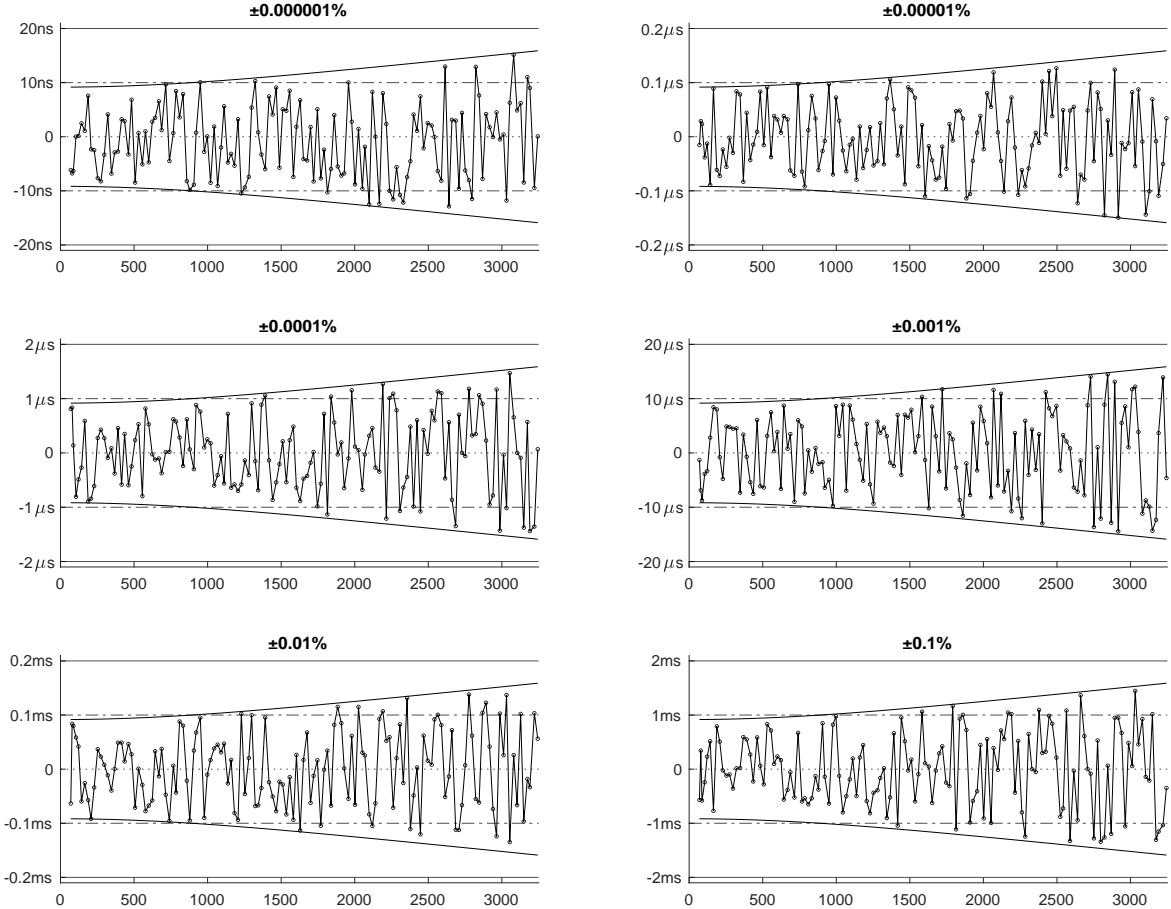}
	\caption{\small%
		Six noise profiles for increasing percentage ranges of traveltimes
	}
	\label{fig:NoiseProfiles}
\end{figure}

Second, we assess the reliability of results under the influence of random noise, which serves to simulate errors in the traveltime data.
The range of the random noise is set at a constant percentage of the measured traveltimes used in the control experiment.
In other words, as the traveltime increases with offset, the range of the noise increases by a fixed percentage.
Within the fixed-percentage range, we generate the noise profile using Matlab's {\tt rand} function, which returns uniformly distributed random numbers within a determined interval.
We generate six noise profiles for which the fixed-percentage range increases by a factor of 10 from $\pm0.000001\%$ to $\pm0.1\%$\,.
As the range increases for each noise profile, the magnitude of noise within the noise profile also increases.
The corresponding mode of the orders of magnitude increases by a factor of 10 from nanoseconds $(10^{-9})$ to milliseconds $(10^{-3})$\,.
Throughout the assessment, we use the same six noise profiles that are plotted in Figure~\ref{fig:NoiseProfiles}.
Similarly to the control experiment, we quantify the goodness of fit of estimator~$\widehat{\boldsymbol{\beta}}$\,.

Third, we perform the control and noise-assessment experiments for a one- and two-layer model.
This process is repeated on eight sets of model parameters, with increasing anisotropy.
We distinguish one set of model parameters from another using notation $[\boldsymbol{\beta}]_s$ for $s\in\{1\,,\dots,8\}$ and their corresponding estimators by $[\widehat{\boldsymbol{\beta}}]_s$\,.
For the one-layer model, we have 
\begin{subequations}
	\label{eq:beta_1_4}
	\begin{gather}
		\underbrace{[\boldsymbol{\beta}]_1 = \left[\mkern-5mu\begin{array}{c}1500\\0.75\\0.0015\end{array}\mkern-5mu\right]}_{\text{near isotropy}}
		\,,\quad
		\underbrace{[\boldsymbol{\beta}]_2 = \left[\mkern-5mu\begin{array}{c}1500\\0.75\\0.0408\end{array}\mkern-5mu\right]}_{\text{low anisotropy}}
		\,,\quad
		\underbrace{[\boldsymbol{\beta}]_3 = \left[\mkern-5mu\begin{array}{c}1500\\0.75\\0.0832\end{array}\mkern-5mu\right]}_{\text{moderate anisotropy}}
		\,,\quad
		\underbrace{[\boldsymbol{\beta}]_4 = \left[\mkern-5mu\begin{array}{c}1500\\0.75\\0.1728\end{array}\mkern-5mu\right]}_{\text{near isotropy}}
		\,;
		\tag{\theequation a,b,c,d}
	\end{gather}
\end{subequations}
for the two-layer model, we have
\begin{subequations}
	\label{eq:beta_5_8}
	\begin{gather}
		\underbrace{[\boldsymbol{\beta}]_5 = \left[\mkern-5mu\begin{array}{c} 911\\3285\\1.5\\0.5\\0.0015\\0.0019\end{array}\mkern-5mu\right]}_{\text{near isotropy}}
		\,,\quad
		\underbrace{[\boldsymbol{\beta}]_6 = \left[\mkern-5mu\begin{array}{c} 911\\3285\\1.5\\0.5\\0.0408\\0.0618\end{array}\mkern-5mu\right]}_{\text{low anisotropy}}
		\,,\quad
		\underbrace{[\boldsymbol{\beta}]_7 = \left[\mkern-5mu\begin{array}{c} 911\\3285\\1.5\\0.5\\0.0832\\0.1272\end{array}\mkern-5mu\right]}_{\text{moderate anisotropy}}
		\,,\quad
		\underbrace{[\boldsymbol{\beta}]_8 = \left[\mkern-5mu\begin{array}{c} 911\\3285\\1.5\\0.5\\0.1728\\0.2688\end{array}\mkern-5mu\right]}_{\text{high anisotropy}}
		\,.
		\tag{\theequation a,b,c,d}
	\end{gather}
\end{subequations}
As we recall from equation~\eqref{eq:chi}, the value of anisotropy corresponds to the difference between the horizontal and vertical velocity.
To quantify the amount of anisotropy, we set four categories based on the relative percentage increase in horizontal to vertical velocity.
They are: near isotropy, ranging from 0\% up to 1\%\,; low anisotropy, from 1\% up to 7\%\,; moderate anisotropy, from 7\% up to 15\%\,; and high anisotropy, from 15\% onward.
For model parameter sets~\eqref{eq:beta_1_4} and~\eqref{eq:beta_5_8}, the anisotropy category to which each belongs is indicated by its respective underbrace.
%

We restrict model-parameter sets to such values to remain consistent with reasonable values of anisotropy exhibited in shale~\citep[e.g.,][pp.~531--532]{VernikLiu1997}.
Finally, we tabulate the observed $[\widehat{\boldsymbol{\beta}}]_s$ and their corresponding $\delta([\widehat{\boldsymbol{\beta}}]_s)$ as well as discuss the reliability of results.

\subsection{One-layer optimization}
\label{sec:OneLayOpt}
For the one-layer model, we seek to solve problem~\eqref{eq:P(beta)}, for which 
\begin{subequations}
	\label{eq:b>0,chi>0}
	\begin{gather}
		l_1(b_1) := \frac{1}{1+\exp\left(-10^6\,b_1\right)}
		\quad{\rm and}\quad
		l_2(\chi_1) := \frac{1}{1+\exp\left(-10^6\,\chi_1\right)}
		\,.
		\tag{\theequation a,b}
	\end{gather}
\end{subequations}
To do so, first, we ensure the convergence of the control experiment on model parameters~\eqref{eq:beta_1_4}, for which we require the use of Hessian modifications and logarithmic barrier functions, which are explained in Sections~\ref{sec:OptTech} and~\ref{sec:LogBar}, respectively, and justified in Appendix~\ref{app:N=1}.
Second, we repeat the optimization on model parameters~\eqref{eq:beta_1_4} under the influence of the six noise profiles plotted in Figure~\ref{fig:NoiseProfiles}.
We tabulate the resulting estimating vectors and their relative errors in Table~\ref{tab:OneLayOpt}.
Therein, we observe two prominent trends concerning the relative errors. 

The first trend is the increase in relative error corresponding to the increase in the range of the noise profile.
We observe that for each increase in the order of magnitude of the noise profile, there is an increase by approximately one order of magnitude of the relative errors of the model parameters.
The second trend is the decrease in relative error corresponding to the increase in anisotropy. 
We observe that, for the nearly isotropic case, the relative-error ordering is $\delta(a_1) < \delta(b_1) < \delta(\chi_1)$\,.
In contrast, as the amount of anisotropy increases, the relative error in $\chi_1$ diminishes to approximately the same level in $a_1$\,, with the largest relative error manifesting in $b_1$\,.
Thus, we state that, for the nearly isotropic case, the relative error manifests disproportionately in the anisotropy parameter whereas, for increasing levels of anisotropy, the relative error is spread out across all model parameters. 
We expect this result because, for low anisotropy, small errors in $\chi_1$ greatly affect the modelled traveltimes, which is not the case for high anisotropy.
In other words, small errors in $\chi_1$ greatly affect the modelled traveltimes in low-anisotropy cases, but have less of an effect for increasing levels of anisotropy.

Along with these trends, we determine that the reliability of results is satisfactory for fixed-percentage noise ranges of at least~$\pm0.1\%$\,.
This percentage corresponds to noise-profile values on the order of milliseconds, which is the same order of magnitude associated with errors in picking traveltimes in real-data cases.
Thus, in the context of optimization, applying a one-layer model to traveltime data in regions where the subsurface is comprised predominantly of shale will yield reliable results.

\begin{table}[H]
	\centering
	\footnotesize
	\begin{tabular}{c*{6}{c}}
		\bottomrule
		& $\pm0.000001\%$ & $\pm0.00001\%$ & $\pm0.0001\%$ & $\pm0.001\%$ & $\pm0.01\%$ & $\pm0.1\%$ \\
		\toprule
		$[\widehat{\boldsymbol{\beta}}]_1$ & 
		$\left[\mkern-10mu\begin{array}{c} 1500.00005\\0.74999993\\0.00150001\end{array}\mkern-10mu\right]$ &
		$\left[\mkern-10mu\begin{array}{c} 1499.99990\\0.75000015\\0.00149998\end{array}\mkern-10mu\right]$ &
		$\left[\mkern-10mu\begin{array}{c} 1499.99866 \\ 0.75000198 \\ 0.00149968\end{array}\mkern-10mu\right]$ &
		$\left[\mkern-10mu\begin{array}{c} 1499.98059 \\ 0.75002729 \\ 0.00149619\end{array}\mkern-10mu\right]$ & 
		$\left[\mkern-10mu\begin{array}{c} 1500.12664 \\ 0.74983824 \\ 0.00150612\end{array}\mkern-10mu\right]$ & 
		$\left[\mkern-10mu\begin{array}{c} 1498.25090 \\ 0.75254910 \\ 0.00109744\end{array}\mkern-10mu\right]$ 
		\\[12.5pt]
		$[\widehat{\boldsymbol{\beta}}]_2$ & 
		$\left[\mkern-10mu\begin{array}{c} 1500.00005 \\ 0.74999993 \\ 0.04080001\end{array}\mkern-10mu\right]$ &
		$\left[\mkern-10mu\begin{array}{c} 1499.99988 \\ 0.75000017 \\ 0.04079998\end{array}\mkern-10mu\right]$ &
		$\left[\mkern-10mu\begin{array}{c} 1499.99842 \\ 0.75000231 \\ 0.04079962\end{array}\mkern-10mu\right]$ &
		$\left[\mkern-10mu\begin{array}{c} 1499.97821 \\ 0.75003055 \\ 0.04079562\end{array}\mkern-10mu\right]$ &
		$\left[\mkern-10mu\begin{array}{c} 1500.13668 \\ 0.74982463 \\ 0.04080671\end{array}\mkern-10mu\right]$ &
		$\left[\mkern-10mu\begin{array}{c} 1497.96923 \\ 0.75294017 \\ 0.04032501\end{array}\mkern-10mu\right]$ 
		\\[12.5pt]
		$[\widehat{\boldsymbol{\beta}}]_3$ & 
		$\left[\mkern-10mu\begin{array}{c} 1500.00006 \\ 0.74999992 \\ 0.08320001\end{array}\mkern-10mu\right]$ &
		$\left[\mkern-10mu\begin{array}{c} 1499.99987 \\ 0.75000020 \\ 0.08319998\end{array}\mkern-10mu\right]$ &
		$\left[\mkern-10mu\begin{array}{c} 1499.99815 \\ 0.75000269 \\ 0.08319955\end{array}\mkern-10mu\right]$ &
		$\left[\mkern-10mu\begin{array}{c} 1499.97551 \\ 0.75003424 \\ 0.08319496\end{array}\mkern-10mu\right]$ &
		$\left[\mkern-10mu\begin{array}{c} 1500.14766 \\ 0.74980972 \\ 0.08320740\end{array}\mkern-10mu\right]$ &
		$\left[\mkern-10mu\begin{array}{c} 1497.64443 \\ 0.75339073 \\ 0.08263901\end{array}\mkern-10mu\right]$ 
		\\[12.5pt]
		$[\widehat{\boldsymbol{\beta}}]_4$ & 
		$\left[\mkern-10mu\begin{array}{c} 1500.00008 \\ 0.74999989 \\ 0.17280002\end{array}\mkern-10mu\right]$ &
		$\left[\mkern-10mu\begin{array}{c} 1499.99983 \\ 0.75000025 \\ 0.17279997\end{array}\mkern-10mu\right]$ &
		$\left[\mkern-10mu\begin{array}{c} 1499.99749 \\ 0.75000360 \\ 0.17279937\end{array}\mkern-10mu\right]$ & 
		$\left[\mkern-10mu\begin{array}{c} 1499.96937 \\ 0.75004267 \\ 0.17279337\end{array}\mkern-10mu\right]$ & 
		$\left[\mkern-10mu\begin{array}{c} 1500.17137 \\ 0.74977744 \\ 0.17280899\end{array}\mkern-10mu\right]$ &
		$\left[\mkern-10mu\begin{array}{c} 1496.88716 \\  0.75443983 \\ 0.17202903\end{array}\mkern-10mu\right]$ \\
		\bottomrule
		& $\pm0.000001\%$ & $\pm0.00001\%$ & $\pm0.0001\%$ & $\pm0.001\%$ & $\pm0.01\%$ & $\pm0.1\%$ \\
		\toprule
		$\delta([\widehat{\boldsymbol{\beta}}]_1)$ & 
		$\left[\mkern-10mu\begin{array}{r} 0.000003\\-0.000009\\0.000589\end{array}\mkern-5mu\right]$ &
		$\left[\mkern-10mu\begin{array}{r} -0.000007\\0.000020\\-0.001186\end{array}\mkern-5mu\right]$ & 
		$\left[\mkern-10mu\begin{array}{r} -0.000089 \\ 0.000264 \\ -0.021257\end{array}\mkern-5mu\right]$ & 
		$\left[\mkern-10mu\begin{array}{r} -0.001294\\ 0.003639 \\ -0.254192\end{array}\mkern-5mu\right]$ & 
		$\left[\mkern-10mu\begin{array}{r} 0.008443 \\ -0.021569 \\ 0.407760\end{array}\mkern-5mu\right]$ & 
		$\left[\mkern-10mu\begin{array}{r} -0.116607 \\ 0.339880 \\ -26.83716\end{array}\mkern-5mu\right]$ 
		\\[12.5pt]
		$\delta([\widehat{\boldsymbol{\beta}}]_2)$ & 
		$\left[\mkern-10mu\begin{array}{r}   0.000004 \\ -0.000010 \\ 0.000025\end{array}\mkern-5mu\right]$ &
		$\left[\mkern-10mu\begin{array}{r} -0.000008 \\ 0.000023 \\ -0.000051\end{array}\mkern-5mu\right]$ &
		$\left[\mkern-10mu\begin{array}{r}  -0.000105 \\ 0.000308 \\ -0.000931\end{array}\mkern-5mu\right]$ &
		$\left[\mkern-10mu\begin{array}{r}  -0.001453 \\ 0.004073 \\ -0.010727\end{array}\mkern-5mu\right]$ &
		$\left[\mkern-10mu\begin{array}{r}   0.009112 \\ -0.023382 \\ 0.016453\end{array}\mkern-5mu\right]$ &
		$\left[\mkern-10mu\begin{array}{r}  -0.135382 \\ 0.392023 \\ -1.164195\end{array}\mkern-5mu\right]$ 
		\\[12.5pt]
		$\delta([\widehat{\boldsymbol{\beta}}]_3)$ & 
		$\left[\mkern-10mu\begin{array}{r}  0.000004 \\ -0.000011 \\ 0.000015\end{array}\mkern-5mu\right]$ &
		$\left[\mkern-10mu\begin{array}{r}  -0.000009 \\ 0.000026 \\ -0.000030\end{array}\mkern-5mu\right]$ &
		$\left[\mkern-10mu\begin{array}{r} -0.000124 \\ 0.000359 \\ -0.000544\end{array}\mkern-5mu\right]$ &
		$\left[\mkern-10mu\begin{array}{r}  -0.001633 \\ 0.004566 \\ -0.006055\end{array}\mkern-5mu\right]$ &
		$\left[\mkern-10mu\begin{array}{r}  0.009844 \\ -0.025371 \\ 0.008893\end{array}\mkern-5mu\right]$ &
		$\left[\mkern-10mu\begin{array}{r}  -0.157038 \\ 0.452098 \\ -0.674271\end{array}\mkern-5mu\right]$ 
		\\[12.5pt]
		$\delta([\widehat{\boldsymbol{\beta}}]_4)$ & 
		$\left[\mkern-10mu\begin{array}{r}  0.000005 \\ -0.000014 \\ 0.000001\end{array}\mkern-5mu\right]$ & 
		$\left[\mkern-10mu\begin{array}{r} -0.000011 \\ 0.000033 \\ -0.000019\end{array}\mkern-5mu\right]$ & 
		$\left[\mkern-10mu\begin{array}{r} -0.000167 \\ 0.000480 \\ -0.000366\end{array}\mkern-5mu\right]$ & 
		$\left[\mkern-10mu\begin{array}{r} -0.002042 \\ 0.005690 \\ -0.003834\end{array}\mkern-5mu\right]$ & 
		$\left[\mkern-10mu\begin{array}{r} 0.011425 \\ -0.029675 \\ 0.005205\end{array}\mkern-5mu\right]$ &
		$\left[\mkern-10mu\begin{array}{r}  -0.207522 \\ 0.591977 \\ -0.446164\end{array}\mkern-5mu\right]$ 
		\\
		\bottomrule
	\end{tabular}
	\captionsetup{singlelinecheck=off}
	\caption[]{\small%
		Tabulated results of performing the inequality-constrained Newton's method in a one-layer model for $[\boldsymbol{\beta}]_1\,,\dots,[\boldsymbol{\beta}]_4$\,, for increasing ranges of constant-percentage noise profiles.
		}
	\label{tab:OneLayOpt}
\end{table}

\subsection{Two-layer optimization}
\label{sec:TwoLayOpt}
For the two-layer model, we seek to solve problem~\eqref{eq:P(beta)}, for which we require $b_i>0$ and $\chi_i>0$\,, for $i=1,2$\,.
As such, we require logarithmic barrier functions~\eqref{eq:b>0,chi>0}, along with another two for the second layer.
Then, we repeat the control experiment on model parameters~\eqref{eq:beta_5_8}, in the same manner described in Appendix~\ref{app:N=1}.
As with the one-layer case, we repeat the optimization under the influence of the six noise profiles plotted in Figure~\ref{fig:NoiseProfiles} and tabulate the resulting estimating vectors, and their relative errors, in Table~\ref{tab:TwoLayOpt}.
Therein, we observe the same trends concerning the relative errors described in Section~\ref{sec:OneLayOpt}---increasing noise-profile values result in increasing relative errors of model parameters, and increasing anisotropy results in greater error for all model parameters as opposed to anisotropy parameters only.

Pertaining to the relative errors in Table~\ref{tab:TwoLayOpt}, the threshold beyond which estimator vectors $\widehat{\boldsymbol{\beta}}$ cease to approximate the model parameters is between fixed-percentage ranges $\pm0.001\%$ and $\pm0.01\%$\,.
The corresponding modes of the orders of magnitude are microseconds and tens of microseconds, respectively, which are several orders of magnitude less than the one-layer-optimization threshold.
In contrast to the one-layer model, the increase in the number of model parameters from three to six reduces drastically the reliability of results for increasing ranges of constant-percentage noise profiles.
With reference to optimization, this is a substantial limitation in applying a two-layer model to real traveltime data.

Let us go further and consider a three-layer model; this would increase the number of model parameters to nine.
The increase from a one- to two-layer model reduces the reliability-of-results threshold from greater than the order of milliseconds to the order of microseconds.
We postulate that there would be---at least---a similar threshold reduction upon increasing the model complexity to three layers.
Apart from the fact that nine-parameter optimization may lead to many local minima and nonphysical results, the severe noise restrictions that beset the three-layer model render its optimization nonviable, especially under the influence of the amount of noise that may be present in real datasets, which we exemplify in Section~\ref{sec:RealDataExample}.
Thus, for pragmatic considerations, we do not consider the application of three-layer models to real traveltime data.

%
%
%

\begin{table}[H]
	\centering
	\scriptsize
	\begin{tabular}{c*{6}{c}}
		\bottomrule
		& $\pm0.000001\%$ & $\pm0.00001\%$ & $\pm0.0001\%$ & $\pm0.001\%$ & $\pm0.01\%$ & $\pm0.1\%$ \\
		\toprule
		$[\widehat{\boldsymbol{\beta}}]_5$ & 
		$\left[\mkern-5mu\begin{array}{c} 911.009592\\3284.96820\\1.49998427\\0.50000299\\0.00149613\\0.00190835\end{array}\mkern-5mu\right]$ & 
		$\left[\mkern-5mu\begin{array}{c} 910.944258\\3285.07556\\1.50011945\\0.50001430\\0.00150593\\0.00187726\end{array}\mkern-5mu\right]$ & 
		$\left[\mkern-5mu\begin{array}{c} 910.266256\\3285.44169\\1.50170736\\0.50042727\\0.00150304\\0.00172162\end{array}\mkern-5mu\right]$ & 
		$\left[\mkern-5mu\begin{array}{c} 914.038158\\3273.10431\\1.49545520\\0.50203289\\0.00003179\\0.00493085\end{array}\mkern-5mu\right]$ & 
		$\left[\mkern-5mu\begin{array}{c} 905.603808\\3288.80029\\1.51216872\\0.50609521\\0.00205428\\0.00003065\end{array}\mkern-5mu\right]$ & 
		$\left[\mkern-5mu\begin{array}{c} 1036.74251\\3267.02054\\1.19449282\\0.56582609\\0.02503647\\0.00002969\end{array}\mkern-5mu\right]$ 
		\\
		$[\widehat{\boldsymbol{\beta}}]_6$ &
		$\left[\mkern-5mu\begin{array}{r} 911.012563 \\ 3284.96100 \\ 1.49997859 \\ 0.50000469 \\ 0.04079494 \\ 0.06181132\end{array}\mkern-5mu\right]$ &
		$\left[\mkern-5mu\begin{array}{r} 910.925996 \\ 3285.10399 \\ 1.50015794 \\ 0.50001434 \\ 0.04080930 \\ 0.06176587\end{array}\mkern-5mu\right]$ &
		$\left[\mkern-5mu\begin{array}{r} 910.039068 \\ 3285.71282 \\ 1.50220374 \\ 0.50049698 \\ 0.04082813 \\ 0.06150877\end{array}\mkern-5mu\right]$ &
		$\left[\mkern-5mu\begin{array}{r} 944.406832 \\ 3196.98077 \\ 1.44009001 \\ 0.51342586 \\ 0.02998654 \\ 0.08788399\end{array}\mkern-5mu\right]$ &
		$\left[\mkern-5mu\begin{array}{r} 839.871930\\3514.61164\\1.62708786\\0.47557539\\0.06823943\\0.00003125\end{array}\mkern-5mu\right]$ &
		$\left[\mkern-5mu\begin{array}{r} 977.211703\\3494.49641\\1.2799358\\0.53941197\\0.09791665\\0.00003108\end{array}\mkern-5mu\right]$ 
		\\
		$[\widehat{\boldsymbol{\beta}}]_7$ &
		$\left[\mkern-5mu\begin{array}{r} 911.016309 \\ 3284.95194 \\ 1.49997138 \\ 0.50000715 \\ 0.08319335 \\ 0.12721535\end{array}\mkern-5mu\right]$ &
		$\left[\mkern-5mu\begin{array}{r} 910.902569 \\ 3285.14096 \\ 1.50020731 \\ 0.50001289 \\ 0.08321418 \\ 0.12714955\end{array}\mkern-5mu\right]$ &
		$\left[\mkern-5mu\begin{array}{r} 909.745255 \\ 3286.08401 \\ 1.50284195 \\ 0.50056564 \\ 0.08326923 \\ 0.12673931\end{array}\mkern-5mu\right]$ &
		$\left[\mkern-5mu\begin{array}{r} 951.816141 \\ 3184.03462 \\ 1.42511862 \\ 0.51800472 \\ 0.06992063 \\ 0.16032199\end{array}\mkern-5mu\right]$ &
		$\left[\mkern-5mu\begin{array}{r} 786.378533 \\ 3753.38368 \\ 1.72096002 \\ 0.43061938 \\ 0.14168995 \\ 0.00003173\end{array}\mkern-5mu\right]$ &
		$\left[\mkern-5mu\begin{array}{r} 1032.35162 \\ 3461.13230 \\ 1.15858605 \\ 0.54271667 \\ 0.14797361 \\ 0.06828564\end{array}\mkern-5mu\right]$ 
		\\
		$[\widehat{\boldsymbol{\beta}}]_8$ &
		$\left[\mkern-5mu\begin{array}{r} 911.026614\\3284.92719\\1.49995141\\0.50001523\\0.17278860\\0.26882767\end{array}\mkern-5mu\right]$ &
		$\left[\mkern-5mu\begin{array}{r} 910.832797\\3285.25726\\1.50035339\\0.49999941\\0.17283187\\0.26869219\end{array}\mkern-5mu\right]$ &
		$\left[\mkern-5mu\begin{array}{r} 908.977690\\3287.06432\\1.50451227\\0.50068564\\0.17300230\\0.26780959\end{array}\mkern-5mu\right]$ &
		$\left[\mkern-5mu\begin{array}{r} 971.071974\\3151.18290\\1.38640710\\0.53100965\\0.15273713\\0.32179400\end{array}\mkern-5mu\right]$ &
		$\left[\mkern-5mu\begin{array}{r} 708.435786\\4260.41204\\1.85699375\\0.26060906\\0.30376789\\0.00003241\end{array}\mkern-5mu\right]$ &
		$\left[\mkern-5mu\begin{array}{r} 1631.52423\\3212.87278\\0.05115720\\0.58844322\\0.23671151\\0.28866875\end{array}\mkern-5mu\right]$ \\
		\bottomrule
		& $\pm0.000001\%$ & $\pm0.00001\%$ & $\pm0.0001\%$ & $\pm0.001\%$ & $\pm0.01\%$ & $\pm0.1\%$ \\
		\toprule
		$\delta([\widehat{\boldsymbol{\beta}}]_5)$ & 
		$\left[\mkern-2.5mu\begin{array}{r} 0.001053\\-0.000968\\-0.001048\\0.000598\\-0.257738\\0.439608\end{array}\mkern-2.5mu\right]$ & 
		$\left[\mkern-2.5mu\begin{array}{r} -0.006119\\0.002300\\0.007963\\0.002859\\0.395071\\-1.197096\end{array}\mkern-2.5mu\right]$ & 
		$\left[\mkern-2.5mu\begin{array}{r} -0.080543\\0.013446\\0.113824\\0.085455\\0.202615\\-9.388697\end{array}\mkern-2.5mu\right]$ & 
		$\left[\mkern-2.5mu\begin{array}{r} 0.333497\\-0.362122\\-0.302987\\0.406579\\-97.88077\\159.5187\end{array}\mkern-2.5mu\right]$ & 
		$\left[\mkern-2.5mu\begin{array}{r} -0.592337\\0.115686\\0.811248\\1.219041\\36.95199\\-98.38681\end{array}\mkern-2.5mu\right]$ & 
		$\left[\mkern-2.5mu\begin{array}{r} 13.80269\\-0.547320\\-20.36715\\13.16522\\1569.098\\-98.43740\end{array}\mkern-2.5mu\right]$ 
		\\
		$\delta([\widehat{\boldsymbol{\beta}}]_6)$ &
		$\left[\mkern-2.5mu\begin{array}{r} 0.001379 \\ -0.001187 \\ -0.001427 \\ 0.000937 \\ -0.012400 \\ 0.018309\end{array}\mkern-2.5mu\right]$ &
		$\left[\mkern-2.5mu\begin{array}{r} -0.008123 \\ 0.003166 \\ 0.010529 \\ 0.002867 \\ 0.022791 \\ -0.055229\end{array}\mkern-2.5mu\right]$ &
		$\left[\mkern-5mu\begin{array}{r} -0.105481 \\ 0.021699 \\ 0.146916 \\ 0.099395 \\ 0.068935 \\ -0.471239\end{array}\mkern-5mu\right]$ &
		$\left[\mkern-2.5mu\begin{array}{r} 3.667051 \\ -2.679429 \\ -3.993999 \\ 2.685172 \\ -26.50357 \\ 42.20711\end{array}\mkern-2.5mu\right]$ &
		$\left[\mkern-2.5mu\begin{array}{r} -7.807692\\6.989700\\8.472524\\-4.884922\\67.25350\\-99.94944\end{array}\mkern-2.5mu\right]$ &
		$\left[\mkern-2.5mu\begin{array}{r} 7.268025\\6.377364\\-14.67095\\7.882394\\139.9918\\-99.94970\end{array}\mkern-2.5mu\right]$ 
		\\
		$\delta([\widehat{\boldsymbol{\beta}}]_7)$ &
		$\left[\mkern-2.5mu\begin{array}{r} 0.001790 \\ -0.001463 \\ -0.001908 \\ 0.001429 \\ -0.007997 \\ 0.012063\end{array}\mkern-2.5mu\right]$ &
		$\left[\mkern-2.5mu\begin{array}{r} -0.010695 \\ 0.004291 \\ 0.013820 \\ 0.002578 \\ 0.017049 \\ -0.039664\end{array}\mkern-2.5mu\right]$ &
		$\left[\mkern-2.5mu\begin{array}{r} -0.137733\\0.032999\\0.189463\\0.113128\\0.083210\\-0.362174\end{array}\mkern-2.5mu\right]$ &
		$\left[\mkern-2.5mu\begin{array}{r} 4.480367 \\ -3.073527 \\ -4.992092 \\ 3.600944 \\ -15.96078 \\ 26.03930\end{array}\mkern-2.5mu\right]$ &
		$\left[\mkern-2.5mu\begin{array}{r} -13.67963 \\ 14.25826 \\ 14.73067 \\ -13.87612 \\ 70.30042 \\ -99.97505\end{array}\mkern-2.5mu\right]$ &
		$\left[\mkern-2.5mu\begin{array}{r} 13.32071 \\ 5.361714 \\ -22.76093 \\ 8.543335 \\ 77.85290 \\ -46.31632\end{array}\mkern-2.5mu\right]$ 
		\\
		$\delta([\widehat{\boldsymbol{\beta}}]_8)$ &
		$\left[\mkern-2.5mu\begin{array}{r} 0.002921\\-0.002216\\-0.003239\\0.003047\\-0.006598\\0.010294\end{array}\mkern-2.5mu\right]$ &
		$\left[\mkern-2.5mu\begin{array}{r} -0.018354\\0.007831\\0.023560\\-0.000118\\0.018446\\-0.040109\end{array}\mkern-2.5mu\right]$ &
		$\left[\mkern-2.5mu\begin{array}{r} -0.221988\\0.062841\\0.300818\\0.137128\\0.117069\\-0.368455\end{array}\mkern-2.5mu\right]$ &
		$\left[\mkern-2.5mu\begin{array}{r} 6.594070\\-4.073580\\-7.572860\\6.201931\\-11.61045\\19.71503\end{array}\mkern-2.5mu\right]$ & 
		$\left[\mkern-2.5mu\begin{array}{r} -22.23537\\29.69291\\23.79958\\-47.87819\\75.79161\\-99.98794\end{array}\mkern-2.5mu\right]$ &
		$\left[\mkern-2.5mu\begin{array}{r} 79.09157\\-2.195654\\-96.58952\\17.68864\\36.98583\\7.391650\end{array}\mkern-2.5mu\right]$ \\
		\bottomrule
	\end{tabular}
	\captionsetup{singlelinecheck=off}
	\caption[]{\small%
		Tabulated results of performing the inequality-constrained Newton's method in a two-layer model for $[\boldsymbol{\beta}]_5\,,\dots,[\boldsymbol{\beta}]_8$\,, for increasing ranges of constant-percentage noise profiles.
	}
	\label{tab:TwoLayOpt}
\end{table}

\section{Real-data example}
\label{sec:RealDataExample}
Let us perform the optimization on a set of measured traveltimes, $\boldsymbol{T}$\,, obtained in offshore Newfoundland~\citep{Kaderali2009}.
The purpose is to use traveltime model~\eqref{eq:t} to solve problem~\eqref{eq:P(beta)} and, in turn, quantify the anisotropy and inhomogeneity of the subsurface in the region. 

\begin{figure}[h]
	\centering
	\includegraphics[width=0.25\textwidth]{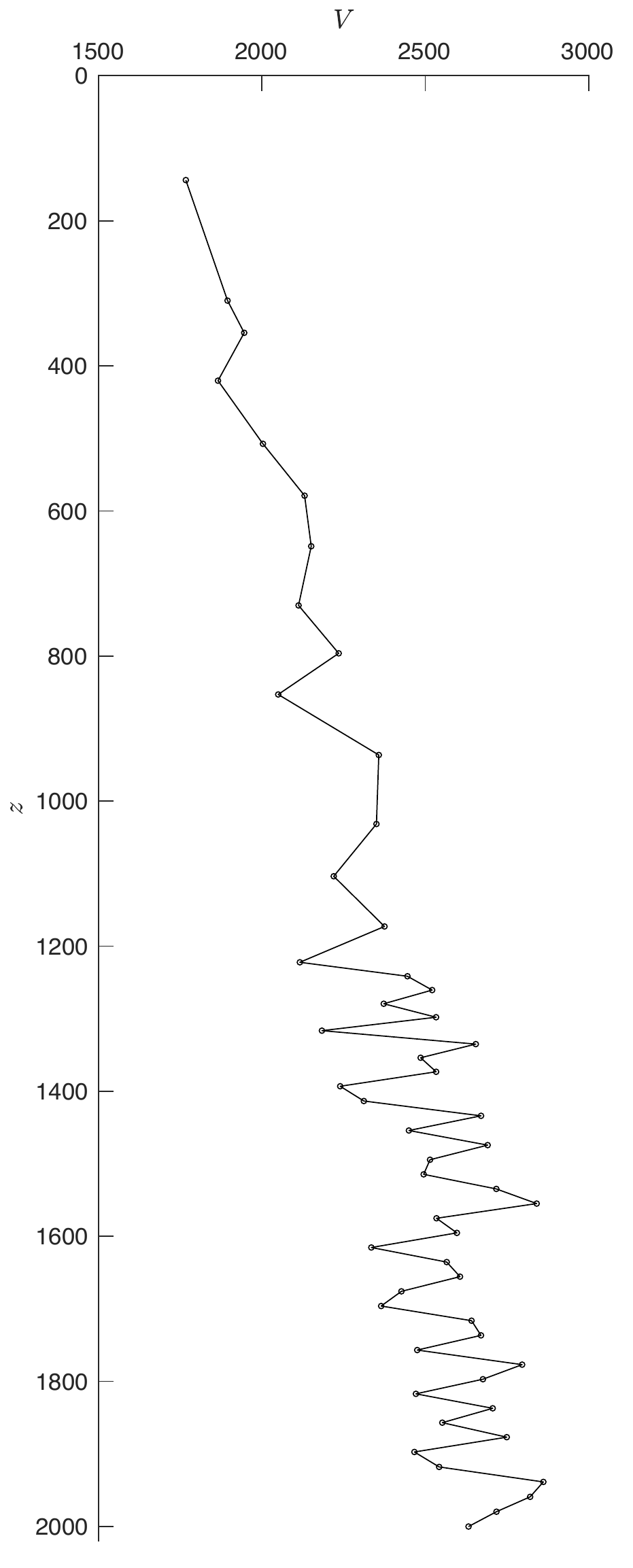}
	\caption{\small Interval velocities with depth}
	\label{fig:Checkshot}
\end{figure}

The subsurface unit in question is comprised predominantly of shale units that can be amalgamated effectively to a single unit.
We are able to corroborate this statement by considering the interval velocities corresponding to the dataset, which we plot in Figure~\ref{fig:Checkshot}.
Therein, we observe a steady linear increase in velocity throughout the entirety of the 2\,km subsurface. 
As such, for this dataset, we consider a one-layer model to be an adequate, albeit low-resolution, representation of the subsurface. 

Further inspection of Figure~\ref{fig:Checkshot} reveals that there are zones of changing interval velocities, namely at depths $z=1212$\,m and $z=1585$\,m\,.
Upon this basis, it is reasonable to apply a multilayer model that segments the subsurface, depending on the desired level of resolution.
Thus, for the two-layer model, we bisect the subsurface and set the layer interface at depth $z_1 = 1212$\,m\,.
We select this layer-interface depth instead of $z=1585$\,m to avoid creating a model with a single layer that dominates the effect of the modelled traveltimes.

\begin{figure}[h]
	\centering
	\includegraphics[width=0.5\textwidth]{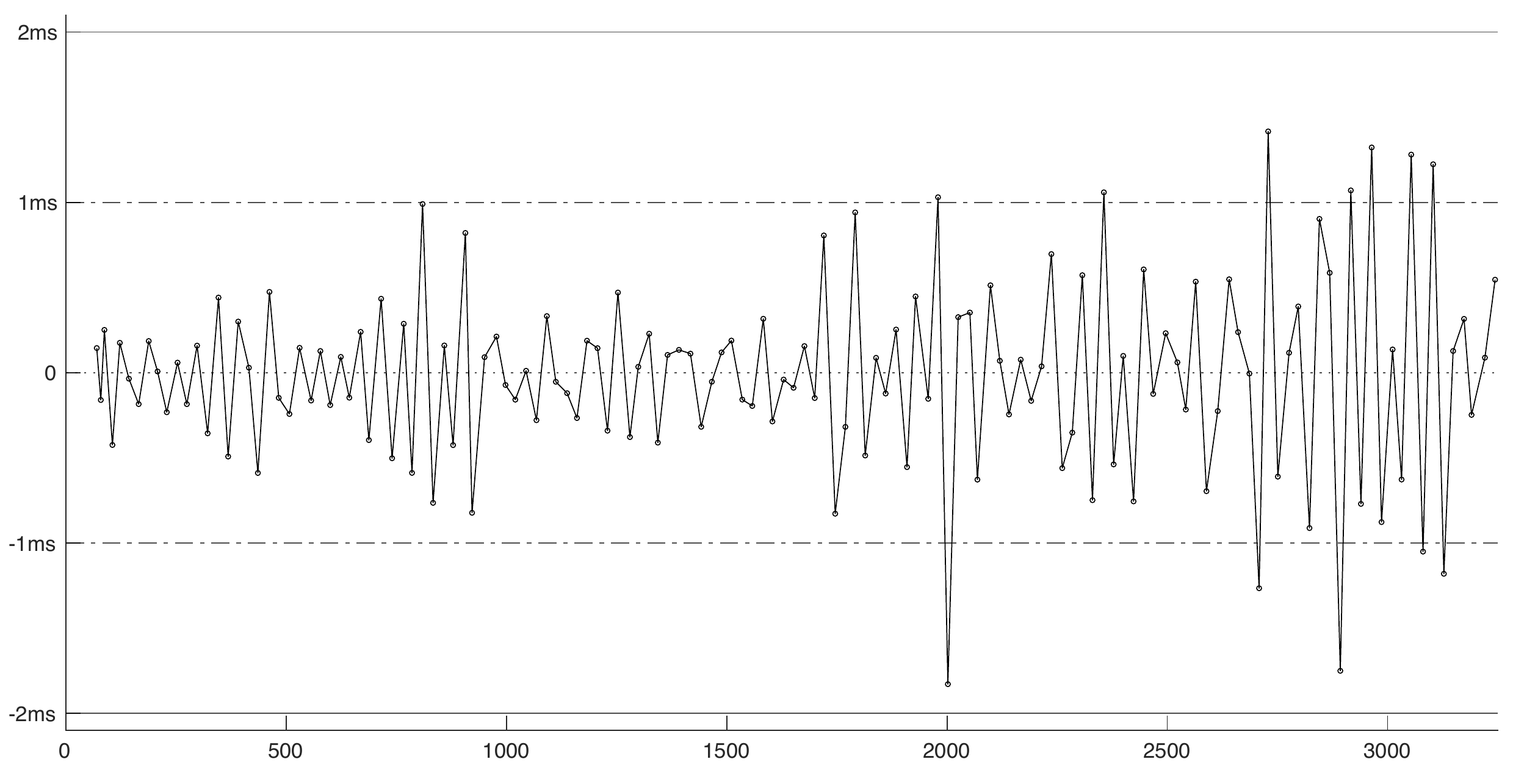}
	\caption{\small Noise profile of measured traveltimes $\boldsymbol{T}$ obtained using a three-point median filter}
	\label{fig:tReal_3pt_NoiseProfile}
\end{figure}

Prior to performing the optimizations, we must assess the amount of noise present in the real data.
To this end, we apply a three-point median filter on the traveltimes and calculate the difference between $\boldsymbol{T}$ and the filtered traveltimes to obtain the noise profile.
We plot the noise profile in Figure~\ref{fig:tReal_3pt_NoiseProfile} and, through visual inspection, we conclude that the amount of noise is on the order of milliseconds.
We calculate that the mode of the orders of magnitude of the noise profile is $10^{-4}$\,, which indicates that the most frequent noise-profile value is on the order of a fraction of a millisecond.

\begin{figure}[h]
	\centering
	\begin{subfigure}[t]{0.49\textwidth}
		\centering
		\includegraphics[width=\textwidth]{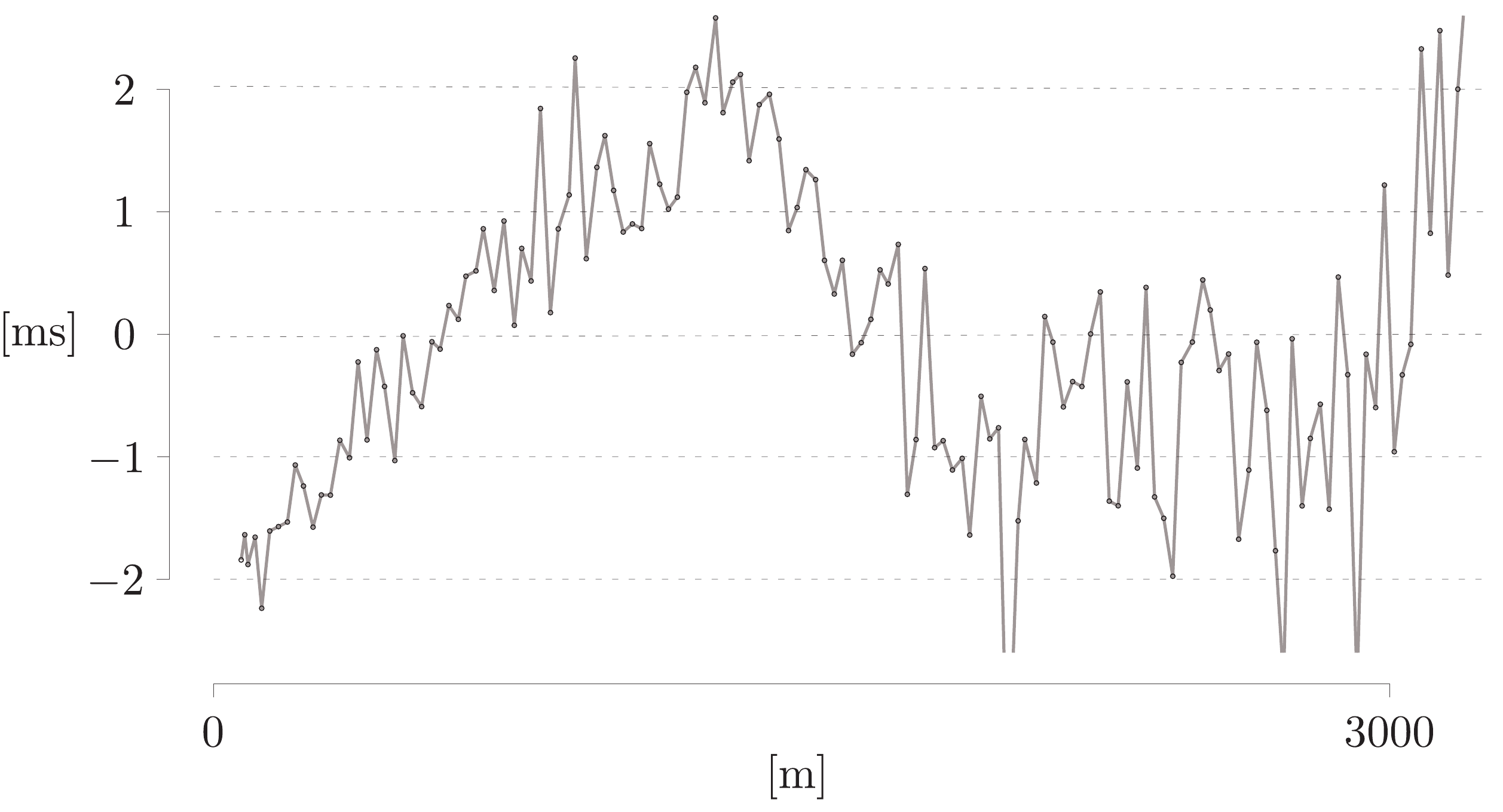}
		\caption{}
		\label{fig:tReal_Res_1Lay}
	\end{subfigure}%
	~
	\begin{subfigure}[t]{0.49\textwidth}
		\centering
		\includegraphics[width=\textwidth]{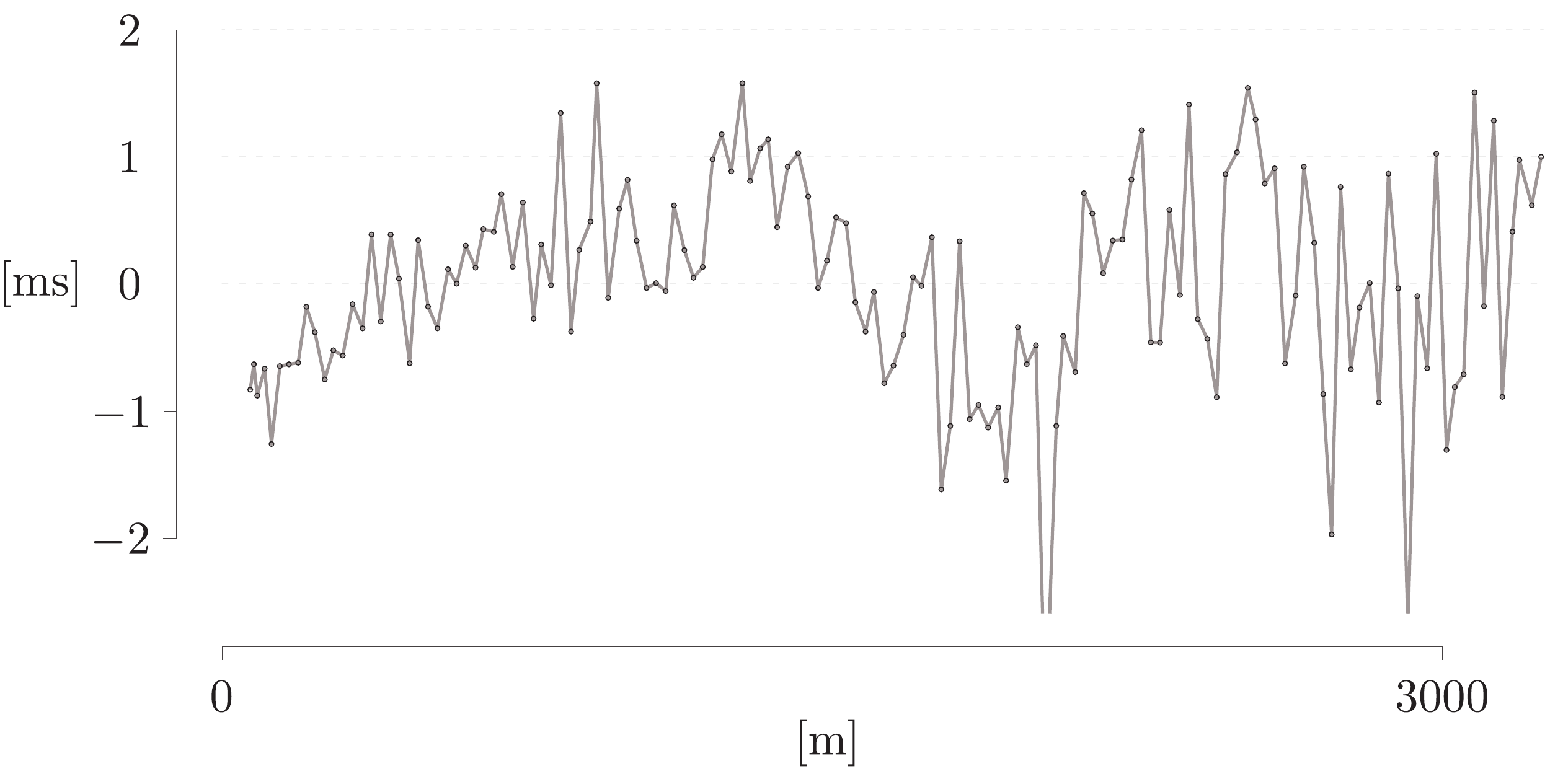}
		\caption{}
		\label{fig:tReal_Res_2Lay}
	\end{subfigure}
	\caption{\small%
		Traveltime residuals with offset upon performing the (a) one- and (b) two-layer optimization on traveltimes~$\boldsymbol{T}$
	}
	\label{fig:tReal_Res}
\end{figure}

As we recall from Sections~\ref{sec:OneLayOpt} and~\ref{sec:TwoLayOpt}, the reliability-of-results threshold for one-layer optimization is greater than milliseconds and for two-layer optimization is on the order of microseconds.
As such, we recognize that, in the context of the optimization, applying a one-layer model to traveltimes $\boldsymbol{T}$ will yield reliable results, but not so for the two-layer model. 
Be that as it may, we perform the optimization for both one- and two-layer models, with restrictions $0<b_i<2$ and $0<\chi_i<1$\,.
The estimator vector and objective-function value, for the former, are
\begin{subequations}
	\label{eq:hat{beta}_N=1,hat{f}_N=1}
	\begin{gather}
		[\widehat{\boldsymbol{\beta}}]_{N=1} = \left[\mkern-2.5mu\begin{array}{r} 1293.79163 \\ 1.12823361 \\ 0.05416071 \end{array}\mkern-2.5mu\right]
		\quad{\rm and}\quad
		\hat{f}_{N=1} = 0.00023430
		\,,
		\tag{\theequation a,b}
	\end{gather}
\end{subequations}
and, for the latter, are
\begin{subequations}
	\label{eq:hat{beta}_N=2,hat{f}_N=2}
	\begin{gather}
		[\widehat{\boldsymbol{\beta}}]_{N=2} = \left[\mkern-2.5mu\begin{array}{r} 897.950408 \\ 3296.42981 \\ 1.98094661 \\ 0.00003242 \\ 0.00002178 \\ 0.00002216\end{array}\mkern-2.5mu\right]
		\quad{\rm and}\quad
		\hat{f}_{N=2} = 0.00009644
		\,.
		\tag{\theequation a,b}
	\end{gather}
\end{subequations}
To gain insight on the values of objective functions~(\ref{eq:hat{beta}_N=1,hat{f}_N=1}b) and~(\ref{eq:hat{beta}_N=2,hat{f}_N=2}b), let us use model parameters~(\ref{eq:hat{beta}_N=1,hat{f}_N=1}a) and~(\ref{eq:hat{beta}_N=2,hat{f}_N=2}a) to calculate the best-fit traveltimes using equation~\eqref{eq:t}.
Taking their difference with traveltimes $\boldsymbol{T}$ for every offset, we obtain the traveltime residuals for  both optimizations, which we plot in Figure~\ref{fig:tReal_Res}.
Therein, we observe that both traveltime residuals vary with offset.
Since the general trends of both the residuals remain within a $\pm$2\,ms range, we maintain that both fits are sufficiently within the traveltime picking error.
However, it is clear that both residual plots exhibit a cyclic pattern, with Figure~\ref{fig:tReal_Res_1Lay}, the one-layer case, being more pronounced than Figure~\ref{fig:tReal_Res_2Lay}, the two-layer case.
We attribute the cyclic pattern to an inadequacy of the model to simulate the measured traveltimes.
The inadequacy may be rooted in an insufficient number of layers to model the subsurface or an absence of elliptical velocity dependence in all layers. 
Within the context of optimization, we expect the residuals to be random and appear to be centred about zero, which is the case of the noise profile of the measured traveltimes in Figure~\ref{fig:tReal_3pt_NoiseProfile}.
To this end, increasing the number of layers not only improves the representation of the medium, but also reduces the cyclic pattern, as can be observed in Figure~\ref{fig:tReal_Res_2Lay}.
With that being said, the cyclic pattern persists, which could be the consequence of imposing elliptical anisotropy in all layers.
Let us examine Figures~\ref{fig:tReal_Res_1Lay} and~\ref{fig:tReal_Res_2Lay} individually.

\captionsetup[subfigure]{subrefformat=simple,labelformat=simple}
\renewcommand\thesubfigure{(\alph{subfigure})}

The traveltime residual in Figure~\ref{fig:tReal_Res_1Lay} does not remain near zero with any offset, which could be attributed to a suboptimal model choice.
Indeed, consolidating nearly two kilometres of subsurface into a single layer does not necessarily lead to an accurate representation of its anisotropy and inhomogeneity.
However, since the subsurface is considered to be predominantly shale and the general trend of the traveltime residual remains within the experimental picking error limits, we claim that this model is an adequate representation of the subsurface, at least at a macroscopic level.
Furthermore, in view of the one-layer-optimization threshold described in Section~\ref{sec:OneLayOpt}, the amount of noise exhibited in $\boldsymbol{T}$ should not lead to poor numerical results.
Thus, we contend that model parameters~(\ref{eq:hat{beta}_N=1,hat{f}_N=1}a) are reliable. 

Recalling the adage that an increase in the number of model parameters results in an increase of numerical fit, the traveltime residual in Figure~\ref{fig:tReal_Res_2Lay} remains closer to zero than in Figure~\ref{fig:tReal_Res_1Lay}, and hence, $\hat{f}_{N=2} < \hat{f}_{N=1}$\,, as expected.
Indeed, bisecting the subsurface improves the numerical fit of traveltimes $\boldsymbol{T}$\,.
However, in view of the reliability-of-results threshold for two-layer optimization described in Section~\ref{sec:TwoLayOpt}, the amount of noise exhibited in $\boldsymbol{T}$\,, which is on the order of milliseconds, is three orders of magnitude greater than the threshold, which is on the order of microseconds.
As such, the reliability of model parameters~(\ref{eq:hat{beta}_N=1,hat{f}_N=1}b) is poor; they indicate that the first layer is strongly inhomogeneous and exhibits low anisotropy, and that the second layer is nearly homogeneous and exhibits near isotropy.
Such results are not consistent with the interval velocities plotted in Figure~\ref{fig:Checkshot}, nor with the knowledge that the subsurface in question is comprised predominantly of shale.
Thus, despite the increase of model adequacy for physical representation of the subsurface and the numerical fit of traveltimes, there is a significant tradeoff in the reliability of results.

\section{Conclusions}
To estimate the values of parameters, attributed to anisotropy and inhomogeneity within the context of an analytical mathematical model, using vertical seismic profile traveltimes, we use Newton's method to minimize the least-squares residual between measured and modelled traveltimes.
To understand applications of this method to real traveltimes, we construct synthetic models to analyze the reliability of results under the influence of increasing percentage ranges of noise.
We determine that there is a threshold for the acceptable level of noise, which increases in severity as the model complexity increases.
In particular, the reliability-of-results threshold for a one-layer model is three orders of magnitude greater than that of a two-layer model.

Within the context of traveltime equation~\eqref{eq:t}, greater complexity is tantamount to segmenting the subsurface into a greater number of layers.
Insofar as constructing physically representative models of the subsurface is concerned, the greater the model complexity, the better the fit, as expected due to the larger number of parameters used to accommodate the data.
However, as determined in Section~\ref{sec:RealDataExample}, the amount of noise present in real traveltime data imposes a tradeoff in the reliability of results with an increase in model complexity.

Pertaining to the real-data example considered in Section~\ref{sec:RealDataExample}, we find that, despite our desire to construct a multilayer model, the noise present in this dataset exceeds the reliability-of-results threshold  for a two-layer model. 
Although we are able to minimize the traveltime residual for a model comprised of two or more layers, we are restricted to a one-layer model for reliable results.
However, since the subsurface in question is comprised predominantly of shale, we contend that the one-layer model is an adequate representation and that model parameters~\eqref{eq:hat{beta}_N=1,hat{f}_N=1} are reliable and acceptable.
\section*{Acknowledgments}
We wish to acknowledge discussions with Michael A. Slawinski and the graphic support of Elena Patarini.
Also, this research was partially supported by the Natural Sciences and Engineering Research Council of Canada, grant 202259.
\bibliographystyle{apa}
\bibliography{KS}

\begin{thebibliography}{}

\bibitem[\protect\astroncite{Aki and Richards}{2002}]{AkiRichards2002}
Aki, K. and Richards, P.~G. (2002).
\newblock {\em Quantitative seismology}, volume~2.
\newblock University Science Books.

\bibitem[\protect\astroncite{{\v C}erven{\'y}}{1989}]{Cerveny1989}
{\v C}erven{\'y}, V. (1989).
\newblock Ray tracing in factorized anisotropic inhomogeneous media.
\newblock {\em Geophysical Journal International}, 99(1):91--100.

\bibitem[\protect\astroncite{{\v C}erven{\'y}}{2001}]{Cerveny2001}
{\v C}erven{\'y}, V. (2001).
\newblock {\em Seismic ray theory}.
\newblock Cambridge University Press.

\bibitem[\protect\astroncite{Danek and Slawinski}{2012}]{DanekSlawinski2012}
Danek, T. and Slawinski, M.~A. (2012).
\newblock Bayesian inversion of {VSP} traveltimes for linear inhomogeneity and
  elliptical anisotropy.
\newblock {\em Geophysics}, 77(6):R239--R243.

\bibitem[\protect\astroncite{Dix}{1955}]{Dix1955}
Dix, C.~H. (1955).
\newblock Seismic velocities from surface measurements.
\newblock {\em Geophysics}, 20(1):68--86.

\bibitem[\protect\astroncite{Fletcher}{2000}]{Fletcher2000}
Fletcher, R. (2000).
\newblock {\em Practical Methods of Optimization}.
\newblock John Wiley \& Sons, Ltd, 2 edition.

\bibitem[\protect\astroncite{Gierlach and Danek}{2018}]{GierlachDanek2018}
Gierlach, B. and Danek, T. (2018).
\newblock Inversion of velocity parameters in multilayered elliptical
  anisotropy medium - synthetic data example.
\newblock {\em E3S Web of Conferences}, 66(01017):1--8.

\bibitem[\protect\astroncite{Helbig}{1994}]{Helbig1994}
Helbig, K. (1994).
\newblock Foundations of anisotropy for exploration seismics.
\newblock In {\em Handbook of geophysical exploration: {S}eismic exploration},
  volume~22. Elsevier Science Serials.

\bibitem[\protect\astroncite{Kaderali}{2009}]{Kaderali2009}
Kaderali, A. (2009).
\newblock Investigating anisotropy and inhomogeneity using tomographic
  inversion of {VSP} traveltimes: Validation of analytic expressions for
  linearly inhomogeneous elliptically anisotropic models.
\newblock Master's thesis, Memorial University of Newfoundland.

\bibitem[\protect\astroncite{Mao and Stuart}{1997}]{MaoStuart1997}
Mao, W. and Stuart, G.~W. (1997).
\newblock Transmission-reflection tomography: {A}pplication to reverse {VSP}
  data.
\newblock {\em Geophysics}, 62(3):884--894.

\bibitem[\protect\astroncite{Nocedal and Wright}{1999}]{NocedalWright1999}
Nocedal, J. and Wright, S.~J. (1999).
\newblock {\em Numerical optimization}.
\newblock Springer series in operations research. Springer.

\bibitem[\protect\astroncite{Pujol et~al.}{1985}]{PujolEtAl1985}
Pujol, J., Burridge, R., and Smithson, S.~B. (1985).
\newblock Velocity determination from offset vertical seismic profiling data.
\newblock {\em Journal of Geophysical Research}, 90(B2):1871--1880.

\bibitem[\protect\astroncite{Rogister and
  Slawinski}{2005}]{RogisterSlawinski2005}
Rogister, Y. and Slawinski, M.~A. (2005).
\newblock Analytic solution of ray-tracing equations for a linearly
  inhomogeneous and elliptically anisotropic velocity model.
\newblock {\em Geophysics}, 70(5):D37--D41.

\bibitem[\protect\astroncite{Slawinski}{2020a}]{Slawinski2020a}
Slawinski, M.~A. (2020a).
\newblock {\em Waves and rays in elastic continua}.
\newblock World Scientific, 4 edition.

\bibitem[\protect\astroncite{Slawinski}{2020b}]{Slawinski2020b}
Slawinski, M.~A. (2020b).
\newblock {\em Waves and rays in seismology: {A}nswers to unasked questions}.
\newblock World Scientific, 3 edition.

\bibitem[\protect\astroncite{Slawinski et~al.}{2004}]{SlawinskiEtAl2004}
Slawinski, M.~A., Wheaton, C.~J., and Powojowski, M. (2004).
\newblock {VSP} traveltime inversion for linear inhomogeneity and elliptical
  anisotropy.
\newblock {\em Geophysics}, 69(2):373--377.

\bibitem[\protect\astroncite{Stewart}{1984}]{Stewart1984}
Stewart, R.~R. (1984).
\newblock {VSP} interval velocities from traveltime inversion.
\newblock {\em Geophysical Prospecting}, 32:608--628.

\bibitem[\protect\astroncite{Thomsen}{1986}]{Thomsen1986}
Thomsen, L. (1986).
\newblock Weak elastic anisotropy.
\newblock {\em Geophysics}, 51(10):1954--1966.

\bibitem[\protect\astroncite{Vernik and Liu}{1997}]{VernikLiu1997}
Vernik, L. and Liu, X. (1997).
\newblock Velocity anisotropy in shales: {A} petrophysical study.
\newblock {\em Geophysics}, 62(2):521--532.

\bibitem[\protect\astroncite{Winterstein}{1990}]{Winterstein1990}
Winterstein, D.~F. (1990).
\newblock Velocity anisotropy terminology for geophysicists.
\newblock {\em Geophysics}, 55(8):956--1112.

\bibitem[\protect\astroncite{Zhang et~al.}{2017}]{ZhangEtAl2017}
Zhang, M., Xu, T., Bai, Z., Liu, Y., Hou, J., and Yu, G. (2017).
\newblock Ray tracing of turning wave in elliptically anisotropic media with an
  irregular surface.
\newblock {\em Earthquake Science}, 30:219--228.

\end{thebibliography}
\begin{appendix}
\setcounter{equation}{0}
\setcounter{table}{0}
\setcounter{figure}{0}
\renewcommand{\theequation}{\Alph{section}.\arabic{equation}}
\renewcommand{\thetable}{\Alph{section}.\arabic{table}}
\renewcommand{\thefigure}{\Alph{section}.\arabic{figure}}
\section{Optimization demonstration for $\boldsymbol{N=1}$}
\label{app:N=1}
In this appendix, we exemplify the optimization technique, described in Section~\ref{sec:OptTech}, used in the control experiment for the one-layer case as well as the justification of the logarithmic barrier function, described in Section~\ref{sec:LogBar}.
\subsection{Control experiment}
\label{app:Opt_N=1}
\begin{figure}[h]
	\centering
	\includegraphics[width=0.6\textwidth]{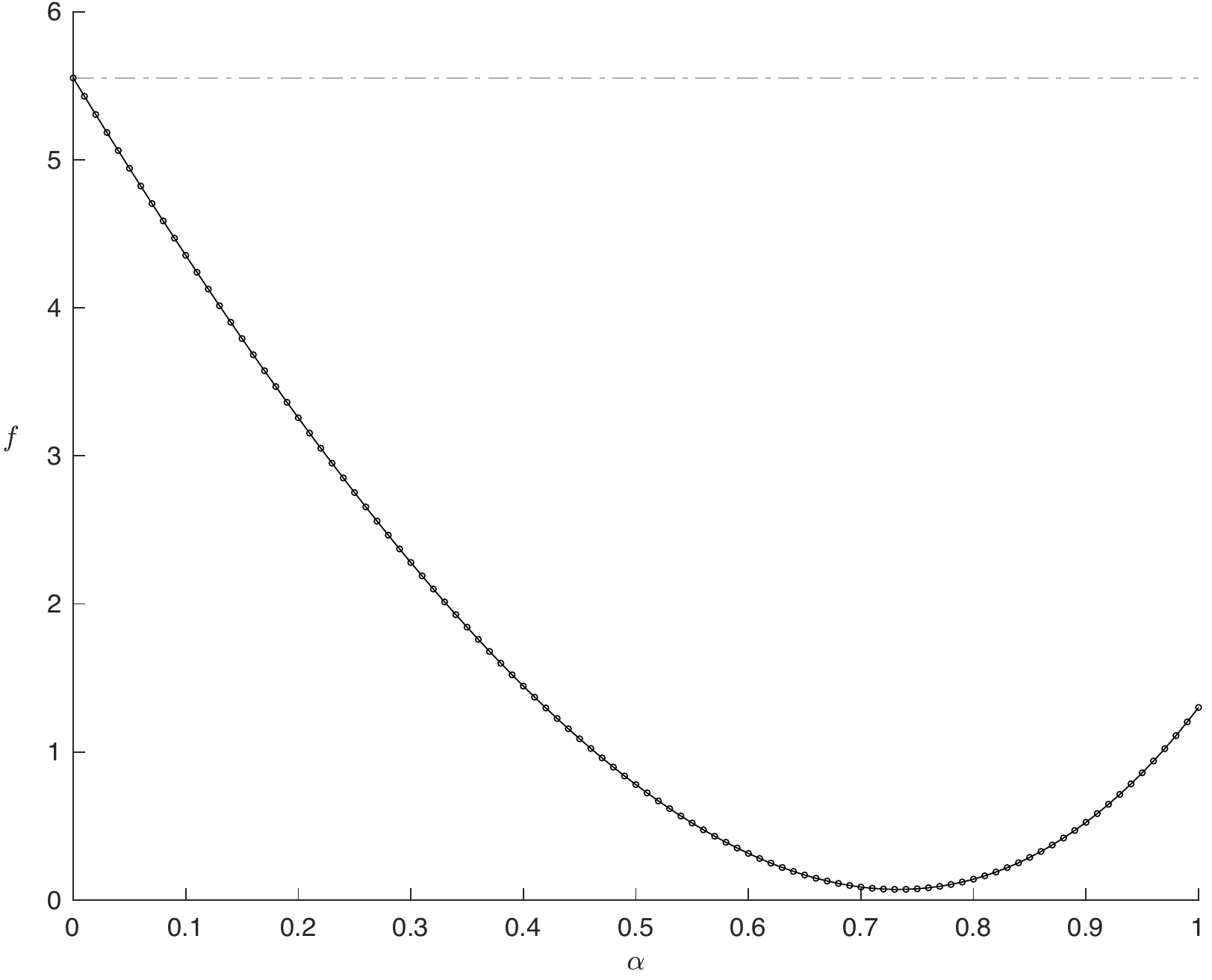}
	\caption{Evaluations of $f(\boldsymbol{\beta}^{(1)} + \alpha\,\Delta\boldsymbol{\beta}^{(1)})$\,, where $\Delta\boldsymbol{\beta}^{(1)} = -\left({\boldsymbol{H}^{\dagger}}^{(1)}\right)^{-1}\boldsymbol{g}^{(1)}$ and $\alpha\in[0,1]$}
	\label{fig:DeltaBeta1_alpha1}
\end{figure}

Let us consider the one-layer case for which the true model parameters are $\boldsymbol{\beta} = [1500,0.75,0.0015]^t$\,.
To perform the optimization, we compute sequences $\boldsymbol{S}$ and generate the measured traveltimes, $\boldsymbol{T}=\boldsymbol{t}(\boldsymbol{S};\boldsymbol{\beta})$\,.
We begin the optimization by selecting a set of starting values, $\boldsymbol{\beta}^{(1)} = [1700,1,0.1]^t$\,, after which we calculate directly the values of the objective function, gradient, and Hessian,
\begin{equation}
	f^{(1)}=5.55225227
	\,,\quad
	\boldsymbol{g}^{(1)} 
	= 
	\left[\mkern-5mu\begin{array}{c} 0.02234679 \\ 18.4370840 \\ 24.9494726\end{array}\mkern-5mu\right]
	\,,\quad
	\boldsymbol{H}^{(1)}
	= 
	\left[\mkern-5mu\begin{array}{ccr} 
		0.00002688 & 0.02337750 & 0.04114517 \\
		0.02337750 & 16.9699206 & 30.3450287 \\
		0.04114517 & 30.3450287 & -13.0394538
	\end{array}\mkern-5mu\right]
	\,.
\end{equation}
Since diagonal entry ${H_{3,3}}^{(1)}$ is negative, it is clear that $\boldsymbol{H}^{(1)}$ is not positive definite, which we verify by its eigenvalues, ${\rm eig}(\boldsymbol{H}^{(1)}) = [-0.00000532,35.8173333,-31.8868343]^t$\,.
As stated in Section~\ref{sec:OptTech}, Newton's method requires a positive definite Hessian to produce a sequence of iterates, $\boldsymbol{\beta}^{(k)}$\,, that are globally convergent. 
Hence, we perform the modified Cholesky factorization to alter $\boldsymbol{H}^{(1)}$ such that it is positive definite; we distinguish the modified Hessian by ${}^{\dagger}$\,.
The result of the alteration is
\begin{equation}
	{\boldsymbol{H}^{\dagger}}^{(1)} 
	:= 
	\left[\mkern-5mu\begin{array}{ccr} 
		0.00009976 & 0.02337750 & 0.04114517 \\
		0.02337750 & 30.7360048 & 30.3450287 \\
		0.04114517 & 30.3450287 & 80.9191363
	\end{array}\mkern-5mu\right]
	\,,\quad
	{\rm where}\quad
	{\rm eig}({\boldsymbol{H}^{\dagger}}^{(1)}) 
	= 
	\left[\mkern-5mu\begin{array}{r} 0.00007558 \\ 16.4523579 \\ 95.2028074\end{array}\mkern-5mu\right]
	\,,
\end{equation}
which satisfies the positive definite requirement.
In view of equation~\eqref{eq:DeltaBeta^{(k)}}, we use $\boldsymbol{H}^{\dagger}$ to calculate the step direction,
\begin{equation}
	\Delta\boldsymbol{\beta}^{(1)}
	=
	\underbrace{-\left[\mkern-5mu\begin{array}{rrr} 
		13231.7683 & -5.43303647 & -4.69057983 \\
		-5.43303647 & 0.05389310 & -0.01744760 \\
		-4.69057983 & -0.01744760 & 0.02128597
	\end{array}\mkern-5mu\right]}_{-\left({\boldsymbol{H}^{\dagger}}^{(1)}\right)^{-1}}
	\underbrace{\left[\mkern-5mu\begin{array}{c} 0.02234679 \\ 18.4370840 \\ 24.9494726\end{array}\mkern-5mu\right]}_{\boldsymbol{g}^{(1)}}
	=
	-\left[\mkern-5mu\begin{array}{r} 78.4907572 \\ 0.43691234 \\ 0.10457155\end{array}\mkern-5mu\right]
	\,.
\end{equation}
To ensure a strict decrease in $f$\,, we must calculate an appropriate step size length, $\alpha$\,, along the $\Delta\boldsymbol{\beta}^{(1)}$ direction.
In Figure~\ref{fig:DeltaBeta1_alpha1}, we plot the values of $f(\boldsymbol{\beta}^{(1)} + \alpha\,\Delta\boldsymbol{\beta}^{(1)})$ for $\alpha\in[0,1]$\,.
Therein, we observe that, for a positive definite Hessian, the resulting curve of residuals resembles a parabola.
In ensuring positive definiteness, we find that this smooth and predictable behaviour is preserved.
However, for nonpositive definiteness, and especially in the presence of noise, we found that we lose this smooth and predicable behaviour, which serves as a tacit justification for ensuring the Hessian remains positive definite for all calculations.

Using Algorithm~\ref{algo:LineSearch}\,, we find that the minimum occurs at $\alpha^{(1)} := 0.73296960$\,, for which we set
\begin{equation}
	\boldsymbol{\beta}^{(2)}
	:=
	\boldsymbol{\beta}^{(1)} + \alpha^{(1)}\,\Delta\boldsymbol{\beta}^{(1)}
	=
	\left[\mkern-5mu\begin{array}{c} 1700 \\ 1 \\ 0.01\end{array}\mkern-5mu\right]
	-
	0.73296960
	\left[\mkern-5mu\begin{array}{r} 78.4907572 \\ 0.43691234 \\ 0.10457155\end{array}\mkern-5mu\right]
	=
	\left[\mkern-5mu\begin{array}{r} 1642.46866 \\ 0.67975654 \\ -0.06664777\end{array}\mkern-5mu\right]
	\,,
\end{equation}
after which we repeat the previous steps by calculating directly the values of the objective function, gradient, and Hessian---the objective function value for $\boldsymbol{\beta}^{(2)}$ is $f^{(2)} = 0.07156195$\,.
The process is repeated until either Algorithm~\ref{algo:LineSearch} produces $\alpha^{(k)}$ whose order of magnitude is less than some value or the decrease in residual is less than some value.
The remaining iterations of the optimization are tabulated in Table~\ref{tab:NoBarOpt_beta_1700_1_0.01}.

\begin{table}[h]
	\centering
	\small
	\begin{tabular}{c*{6}{c}}
		\toprule
		$k$ & & \multicolumn{3}{c}{$\{\boldsymbol{\beta}^{(k)}\}$} & & $f^{(k)}$ \\
		\cmidrule{1-1}\cmidrule{3-5}\cmidrule{7-7}
		1 & & 1700 & 1 & 0.01 & & 5.55225227 \\
		2 & & 1642.46866 & 0.67975654 & $-$0.06664777 & & 0.07156195 \\
		3 & & 1303.54209 & 1.07977513 & $-$0.07656328 & & 0.01186727 \\
		4 & & 1346.92235 & 0.98946329 & $-$0.04426630 & & 0.00239513 \\
		5 & & 1436.89201 & 0.84021609 & $-$0.01334487 & & 0.00051080 \\
		6 & & 1473.69567 & 0.78820908 & $-$0.00461155 & & 0.00004540 \\
		7 & & 1497.03543 & 0.75408963 & 0.00091052 & & 0.00000097 \\
		8 & & 1499.90130 & 0.75014074 & 0.00147849 & & 6.3848$\times10^{-10}$ \\
		9 & & 1499.99994 & 0.75000008 & 0.00149999 & & 3.3154$\times10^{-16}$ \\
		10 & & 1500 & 0.75 & 0.0015 & & 2.0268$\times10^{-26}$ \\
		\bottomrule
	\end{tabular}
	\caption{\small
		Iterations of unconstrained Newton's method performed on the one-layer case using Hessian modifications from starting values $\boldsymbol{\beta}^{(1)} = [1700,1,0.01]^t$\,, converging to true values $[\boldsymbol{\beta}]_1 = [1500,0.75,0.0015]^t$\,.}
	\label{tab:NoBarOpt_beta_1700_1_0.01}
\end{table}

\subsection{Control experiment without Hessian modification}
\label{app:Opt_N=1_NoMods}
In Section~\ref{app:Opt_N=1}, we ensured a positive definite Hessian throughout the optimization using the modified Cholesky factorization.
However, if the Hessian is not altered, the optimization is not guaranteed to converge to the true solution from all starting values.
Indeed, considering the same model parameters, $\boldsymbol{\beta}$\,, and starting values, $\boldsymbol{\beta}^{(1)}$\,, the optimization converges to the correct solution.
It is not the case from another set of starting values, say, $\boldsymbol{\beta}^{(1)} = [2400,1,0.2]^t$\,.
To demonstrate this, we calculate the values of the objective function, gradient, and Hessian,
\begin{equation}
	f^{(1)}=31.08341448
	\,,\quad
	\boldsymbol{g}^{(1)} 
	= 
	\left[\mkern-5mu\begin{array}{c} 0.03018866 \\ 25.0184043 \\ 28.8129228\end{array}\mkern-5mu\right]
	\,,\quad
	\boldsymbol{H}^{(1)}
	= 
	\left[\mkern-5mu\begin{array}{rrr} 
		-0.00000424 & -0.00248880 & 0.00556654\\
		-0.00248880 & -4.71975983 & 1.49180974 \\
		 0.00556654 &  1.49180975 & -53.6965777
	\end{array}\mkern-5mu\right]
	\,.
\end{equation}
As indicated by its main-diagonal entries and its eigenvalues, 
\begin{equation}
	{\rm eig}(\boldsymbol{H}^{(1)}) = 
	[-53.7419760, -0.00000250, -4.67436327]^t
	\,,
\end{equation}
the Hessian is negative definite.
Without a modification to its values, we use $\boldsymbol{H}^{(1)}$ to calculate the search direction.
We obtain
\begin{equation}
	\Delta\boldsymbol{\beta}^{(1)}
	= 
	[8112.60299, 1.47122367, 1.41846688]^t
	\,,
\end{equation}
which does not permit a decrease in $f$ for any step size length.
As such, the optimization is terminated on the first iteration.
For other starting values, it is not guaranteed to encounter such a scenario on the first iteration, but the risk remains at each iteration of the optimization.

\begin{table}[h]
	\centering
	\small
	\begin{tabular}{c*{6}{c}}
		\toprule
		$k$ & & \multicolumn{3}{c}{$\{\boldsymbol{\beta}^{(k)}\}$} & & $f^{(k)}$ \\
		\cmidrule{1-1}\cmidrule{3-5}\cmidrule{7-7}
		1 & & 2400 & 1 & 0.2 & & 31.0834145 \\
		2 & & 2204.86452 & $-$0.31361642 & 0.10542578 & & 1.50964783 \\
		3 & & 2327.63552 & $-$0.31084822 & 0.11061909 & & 0.07863261 \\
		4 & & 2389.84780 & $-$0.31372871 & 0.07165287 & & 0.00606885 \\
		5 & & 2497.11162 & $-$0.40068941 & 0.04845627 & & 0.00228533 \\
		6 & & 2678.09091 & $-$0.57165728 & 0.03132588 & & 0.00118914 \\
		7 & & 2792.97026 & $-$0.66992183 & 0.01475213 & & 0.00020636 \\
		8 & & 2873.14100 & $-$0.73880958 & 0.00362632 & & 0.00001206 \\
		9 & & 2885.52025 & $-$0.74884791 & 0.00169315 & & 0.00000004 \\
		10 & & 2886.87975 & $-$0.74999999 & 0.0015 & & 1.1947$\times10^{-22}$ \\
		11 & & 2886.87975 & $-$0.75 & 0.0015 & & 3.0863$\times10^{-26}$  \\
		\bottomrule
	\end{tabular}
	\caption{\small
	Iterations of unconstrained Newton's method performed on the one-layer case using Hessian modifications from starting values $\boldsymbol{\beta}^{(1)} = [2400,1,0.2]^t$\,, not converging to true values $[\boldsymbol{\beta}]_1 = [1500,0.75,0.0015]^t$\,.}
	\label{tab:NoBarOpt_beta_2400_1_0.2}
\end{table}

In view of this result, we insist that the Hessian remains positive definite throughout all executions of the optimization.
Thus, we repeat the optimization for starting values $\boldsymbol{\beta}^{(1)} = [2400,1,0.2]^t$\,, following the procedure indicated in Section~\ref{app:Opt_N=1}.
The iterations of the optimization are tabulated in Table~\ref{tab:NoBarOpt_beta_2400_1_0.2}.
Therein, we observe that the optimization converges to
\begin{equation}
	\label{eq:app:hat{beta}_N=1_nonphysical}
	\widehat{\boldsymbol{\beta}} 
	= 
	\left[\mkern-5mu\begin{array}{c} 2886.87975 \\ -0.75 \\ 0.0015\end{array}\mkern-5mu\right]
	\neq
	\left[\mkern-5mu\begin{array}{c} 1500 \\ 0.75 \\ 0.0015\end{array}\mkern-5mu\right]
	=
	\boldsymbol{\beta}
	\,.
\end{equation}

Under the context of the optimization, estimator~\eqref{eq:app:hat{beta}_N=1_nonphysical} is an acceptable solution.
However, in practice, we reject this solution as we expect $b_i>0$ and $\chi_i>0$ due to the physical considerations indicated in Section~\ref{eq:abchiModel}.
Namely, model parameter $\widehat{\beta}_2 = b_1 = -0.75 < 0$ is a nonphysical result.

\subsection{Logarithmic barrier functions in control experiment}
\label{app:Barrier}
In Section~\ref{app:Opt_N=1_NoMods}, we observed that the optimization used in the control experiment can result in solutions with nonphysical parameters.
To ensure that solutions remain physically meaningful, we use the logarithmic barrier function, described in Section~\ref{sec:LogBar}, to restrict the values of model parameters.
In other words, we introduce logarithmic barrier functions to $f$ that change the unconstrained optimization problem to an inequality-constrained optimization problem.
In view of equation~\eqref{eq:P(beta)}, the objective function for this problem is
\begin{equation}
	P(\boldsymbol{\beta}) = f(\boldsymbol{\beta}) - \sum\limits_{i\in\mathcal{I}}\,\log l_i\,.
\end{equation}

\begin{figure}[h]
	\centering
	\includegraphics[width=0.495\textwidth]{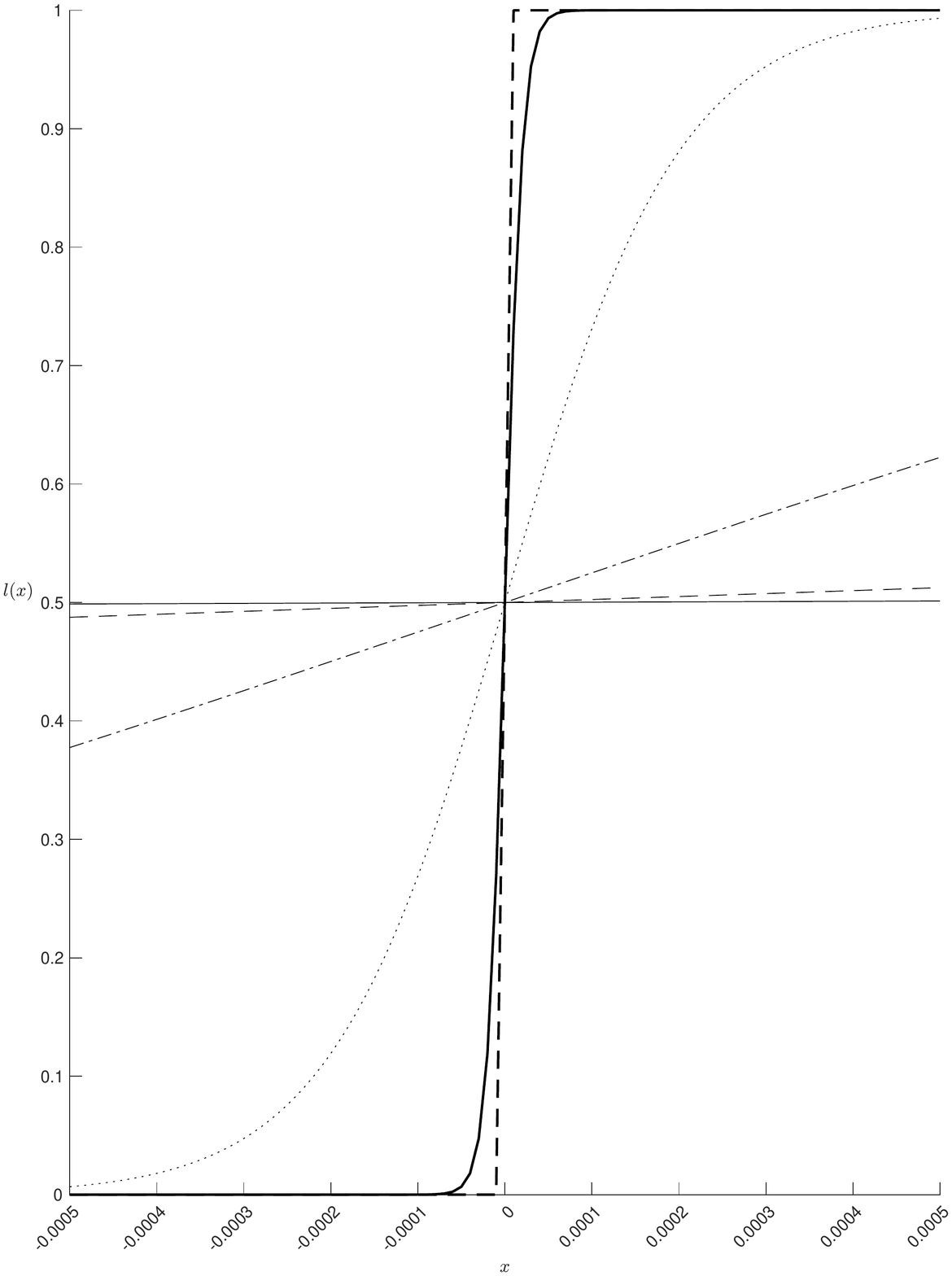}
	\includegraphics[width=0.495\textwidth]{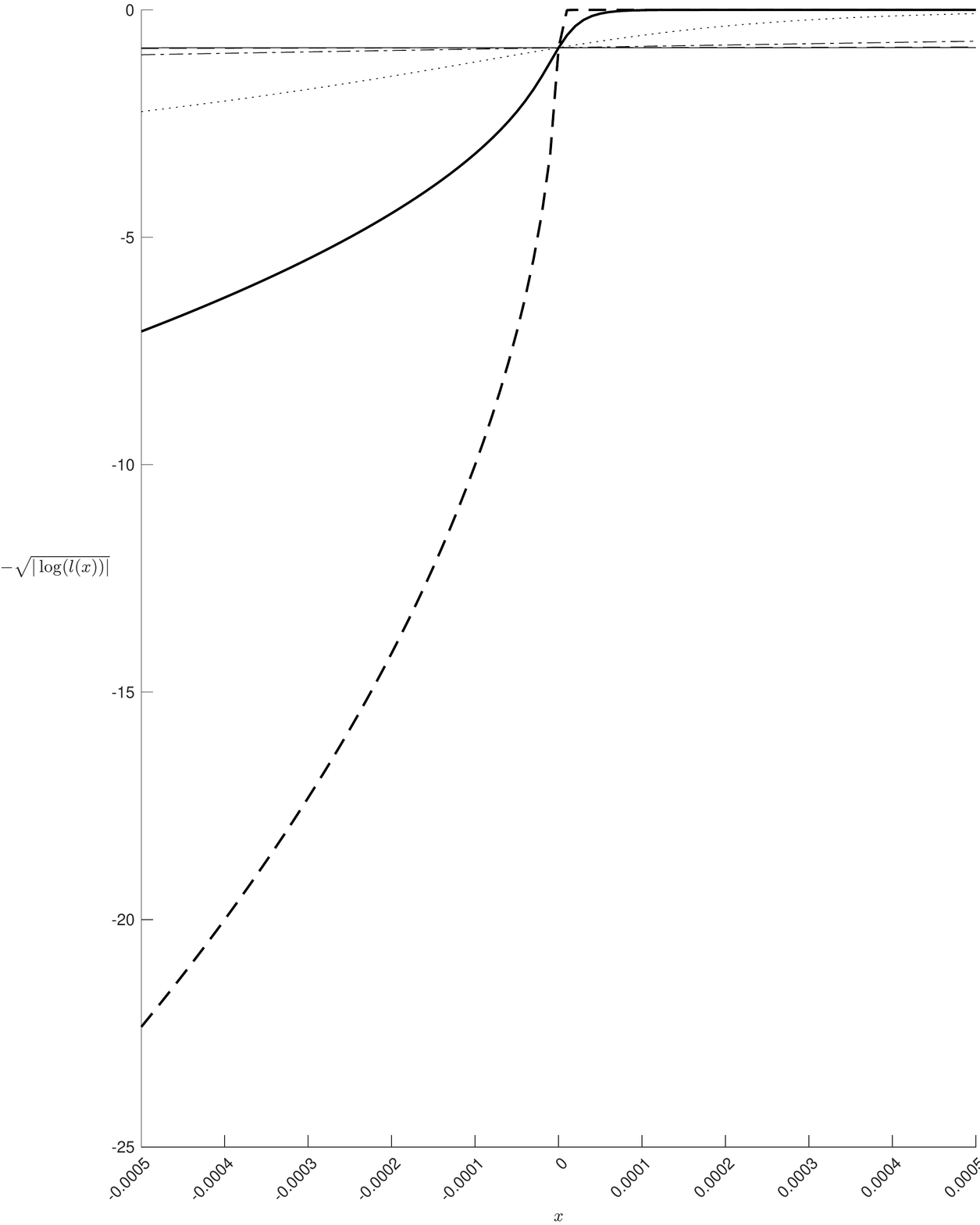}
	\caption{\small%
		Demonstration of increasing growth rate, $r$\,, of logistic function, $l(x)$\,, centred on $x_0 = 0$\,, and the logarithm of the logistic function, which is the logarithmic barrier function.
		We plot the logistic function on the left-hand side.
		However, the barrier function increases too rapidly to be plotted adequately. 
		Hence, we manipulate the vertical scale and plot $-\sqrt{|\log(l(x)|}$ on the right-hand side, which demonstrates adequately the increasing negative barrier-function values. 
		Herein, we increase $r$ by orders of 10 until $l$ resembles a step function for $x$ on the order of $10^{-4}$\,.
		For both plots, the solid line corresponds to $r=10$\,, dashed to $r=10^2$\,, dot-dashed to $r=10^3$\,, dotted to $r=10^4$\,, solid bold to $r=10^5$\,, and dashed bold to $r=10^6$\,.
	}
	\label{fig:PosBar_b1}
\end{figure}

To maintain the restrictions imposed by physical considerations, $b_1>0$ and $\chi_1>0$\,, we recall the logistic function for some argument $x$\,,
\begin{equation}
	\label{eq:app:l}
	l(x) = \frac{1}{1+\exp\left(-r\left(x-x_0\right)\right)}
	\,.
\end{equation}
To use logistic function~\eqref{eq:app:l} as a logarithmic barrier function for $b_1>0$ and $\chi_1>0$\,, we require that it resembles a step function ranging from zero to one with a midpoint of $x = 0$\,, which requires a large growth rate, $r$\,.
We select the value of $r$ by increasing its magnitude and inspecting its curve in the left-hand plot of Figure~\ref{fig:PosBar_b1}.
Since we do not expect $\chi_1<10^{-3}$\,, we require that the value of $r$ that results in a step-like logistic function for $\chi_1$ on the order of $10^{-4}$\,---this value is $r=10^6$\,.
Also, since we expect the order of magnitude of $b_i$ greater than or equal to the order of magnitude of $\chi_i$\,, we use $r=10^6$ for its barrier function as well.

As such, we restate equations~\eqref{eq:b>0,chi>0} so that, for both restrictions, we have
\begin{subequations}
	\label{eq:app:b>0,chi>0}
	\begin{gather}
		l_1(b_1) := \frac{1}{1+\exp\left(-10^6\,b_1\right)}
		\quad{\rm and}\quad
		l_2(\chi_1) := \frac{1}{1+\exp\left(-10^6\,\chi_1\right)}
		\,.
		\tag{\theequation a,b}
	\end{gather}
\end{subequations}
Taking the logarithm of equation~(\ref{eq:app:b>0,chi>0}b) and plotting the curves on the right-hand plot of Figure~\ref{fig:PosBar_b1}, we observe that its value equals zero for $\chi_1>0$\,, and rapidly increases in negative value outside that domain.
We tabulate these values for $\chi_1$ in Table~\ref{tab:barrier_b1}; since the restrictions for $b_1$ and $\chi_1$ are the same, the barrier-function values in Table~\ref{tab:barrier_b1} are the same for both $b_1$ and $\chi_1$\,. 
Subtracting the logarithms of equations~\eqref{eq:app:b>0,chi>0} from $f(\boldsymbol{\beta})$ results in increasingly large positive values of $P$ for $b_1<0$ and $\chi_1<0$\,.
Effectively, this restricts the optimization to positive values $b_1$ and $\chi_1$\,.

\begin{table}[h]
	\small
	\centering
	\begin{tabular}{c*{2}{c}}
		\toprule
		$\chi_1$ & $l_2$ & $\log(l_2)$ \\
		\toprule
		$-$0.0010 & 0 & $-\infty$ \\
		$-$0.0009 & 0 & $-\infty$ \\
		$-$0.0008 & 0 & $-\infty$ \\
		$-$0.0007 & 9.8597$\times10^{-305}$ & $-$700 \\
		$-$0.0006 & 2.6504$\times10^{-261}$ & $-$600 \\
		$-$0.0005 & 7.1246$\times10^{-218}$ & $-$500 \\
		$-$0.0004 & 1.9152$\times10^{-174}$ & $-$400 \\
		$-$0.0003 & 5.1482$\times10^{-131}$ & $-$300 \\
		$-$0.0002 & 1.38390$\times10^{-87}$ & $-$200 \\
		$-$0.0001 & 3.72007$\times10^{-44}$ & $-$100 \\
		0 & 0.5 & $-$0.693 \\
		0.0001 & 1 & 0 \\
		0.0002 & 1 & 0 \\
		0.0003 & 1 & 0 \\
		0.0004 & 1 & 0 \\
		0.0005 & 1 & 0 \\
		0.0006 & 1 & 0 \\
		0.0007 & 1 & 0 \\
		0.0008 & 1 & 0 \\
		0.0009 & 1 & 0 \\
		0.0010 & 1 & 0 \\
		\bottomrule
	\end{tabular}
	\caption{Increasingly negative values of logarithmic barrier function for $\chi_1 < 0$}
	\label{tab:barrier_b1}
\end{table}

Using restrictions~\eqref{eq:app:b>0,chi>0}, we repeat the optimization for starting values $\boldsymbol{\beta}^{(1)} = [1700,1,0.01]^t$ and $\boldsymbol{\beta}^{(1)} = [2400,1,0.2]^t$\,.
The iterations of the optimization are tabulated, for the former, in Table~\ref{tab:PosBar_beta_1700_1_0.01} and, for the latter, in Table~\ref{tab:PosBar_beta_2400_1_0.2}.

Let us compare the sequence of iterates for the unconstrained and inequality-constrained problems.
For the former, the value of $\chi_1$ in the unconstrained problem  becomes negative on the second iteration and improves the residual by nearly an order of magnitude for each of the subsequent four iterations before returning to positive values.
In the inequality-constrained problem, the value of $\chi_1$ attempts to become negative on the second iteration, but is restricted by the logarithmic barrier function.
For the following fourteen iterations, the improvement of the objective function is slow as a consequence of the value of $\chi_1$ being affected by the barrier.
Once the value of $\chi_1$ is unaffected by the barrier, the pace of the optimization improves considerably and it converges to the correct solution within four iterations.

For the latter, the value of $b_1$ in the unconstrained problem becomes negative on the second iteration, which leads the optimization toward a nonphysical solution.
Similarly, in the inequality-constrained problem, the value of $b_1$ attempts to become negative on the second iteration, but is restricted by the logarithmic barrier function.
The progress of the optimization becomes extremely slow as the objective function's order of magnitude does not change for 24 iterations as both $b_1$ and $\chi_1$ appear to undergo marginal improvements due to their proximity to the barriers.
Although the number of iterations nearly triples in comparison to the unconstrained problem, the inequality-constrained optimization converges to the correct solution, which was not possible without the barriers. 

The results of both optimizations demonstrate the necessity of the logarithmic barrier functions to guarantee convergence from different starting points. 
As such, we include such functions in all calculations.

\begin{table}[h]
	\centering
	\small
	\begin{tabular}{c*{6}{c}}
		\toprule
		$k$ & & \multicolumn{3}{c}{$\{\boldsymbol{\beta}^{(k)}\}$} & & $P^{(k)}$ \\
		\cmidrule{1-1}\cmidrule{3-5}\cmidrule{7-7}
		1 & & 1700 & 1 & 0.01 & & 5.55225227 \\
		2 & & 1692.54515 & 0.95850321 & 0.00006807 & & 4.41369483 \\
		3 & & 1233.20128 & 1.08472339 & 0.00007378 & & 0.03674654 \\
		4 & & 1426.63490 & 0.83058675 & 0.00008276 & & 0.00451468 \\
		5 & & 1478.73012 & 0.77649374 & 0.00009275 & & 0.00020029 \\
		6 & & 1494.14148 & 0.75855244 & 0.00010275 & & 0.00003661 \\
		7 & & 1494.76944 & 0.75783063 & 0.00011276 & & 0.00001461 \\
		8 & & 1494.80826 & 0.75777284 & 0.00012278 & & 0.00000656 \\
		9 & & 1494.84601 & 0.75771622 & 0.00013284 & & 0.00000359 \\
		10 & & 1494.88414 & 0.75765904 & 0.00014300 & & 0.00000247 \\
		11 & & 1494.92333 & 0.75760026 & 0.00015345 & & 0.00000205 \\
		12 & & 1494.96553 & 0.75753698 & 0.00016469 & & 0.00000187 \\
		13 & & 1495.01729 & 0.75745936 & 0.00017848 & & 0.00000178 \\
		14 & & 1495.11025 & 0.75731996 & 0.00020324 & & 0.00000170 \\
		15 & & 1495.72385 & 0.75639986 & 0.00036667 & & 0.00000129 \\
		16 & & 1499.97374 & 0.75003000 & 0.00149732 & & 6.7360$\times10^{-10}$ \\
		17 & & 1499.99998 & 0.75000003 & 0.00149999 & & 2.8015$\times10^{-17}$ \\
		18 & & 1500 & 0.74999999 & 0.0015 & & 2.4500$\times10^{-26}$ \\
		19 & & 1500 & 0.75 & 0.0015 & & 9.4772$\times10^{-27}$ \\
 		\bottomrule
	\end{tabular}
	\caption{\small
	Iterations of inequality-constrained Newton's method performed on the one-layer case using Hessian modifications from starting values $\boldsymbol{\beta}^{(1)} = [1700,1,0.01]^t$\,, converging to true values $[\boldsymbol{\beta}]_1 = [1500,0.75,0.0015]^t$\,.}
	\label{tab:PosBar_beta_1700_1_0.01}
\end{table}

\begin{table}[h]
	\centering
	\small
	\begin{tabular}{c*{6}{c}}
		\toprule
		$k$ & & \multicolumn{3}{c}{$\{\boldsymbol{\beta}^{(k)}\}$} & & $P^{(k)}$ \\
		\cmidrule{1-1}\cmidrule{3-5}\cmidrule{7-7}
		1 & & 2400 & 1 & 0.2 & & 31.0834145 \\
		2 & & 2251.46463 & 0.00008704 & 0.12801096 & & 1.61680199 \\
		3 & & 2041.92934 & 0.00009425 & 0.14563548 & & 0.04690324 \\
		4 & & 2091.67981 & 0.00010141 & 0.07345702 & & 0.00623122 \\
		5 & & 2101.53157 & 0.00011141 & 0.06829222 & & 0.00456089 \\
		6 & & 2102.01283 & 0.00012141 & 0.06785146 & & 0.00454966 \\
		7 & & 2102.00406 & 0.00013141 & 0.06785129 & & 0.00454629 \\
		8 & & 2101.99481 & 0.00014141 & 0.06785129 & & 0.00454505 \\
		9 & & 2101.98557 & 0.00015141 & 0.06785129 & & 0.00454459 \\
		10 & & 2101.97632 & 0.00016141 & 0.06785129 & & 0.00454442 \\
		11 & & 2101.96707 & 0.00017142 & 0.06785129 & & 0.00454436 \\
		12 & & 2101.95781 & 0.00018144 & 0.06785129 & & 0.00454434 \\
		13 & & 2101.94852 & 0.00019149 & 0.06785129 & & 0.00454433 \\
		14 & & 2101.93914 & 0.00020163 & 0.06785129 & & 0.00454433 \\
		15 & & 2101.92950 & 0.00021205 & 0.06785129 & & 0.00454433 \\
		16 & & 2101.91910 & 0.00022331 & 0.06785129 & & 0.00454432 \\
		17 & & 2101.90605 & 0.00023742 & 0.06785129 & & 0.00454432 \\
		18 & & 2101.87932 & 0.00026633 & 0.06785128 & & 0.00454432 \\
		19 & & 2100.71385 & 0.00152699 & 0.06785119 & & 0.00454429 \\
		20 & & 2099.30273 & 0.00305433 & 0.06785037 & & 0.00454416 \\
		21 & & 2096.48020 & 0.00611086 & 0.06784760 & & 0.00454368 \\
		22 & & 2090.82795 & 0.01223931 & 0.06783648 & & 0.00454173 \\
		23 & & 2079.43643 & 0.02462133 & 0.06779151 & & 0.00453382 \\
		24 & & 2055.79366 & 0.05044920 & 0.06760345 & & 0.00450040 \\
		25 & & 1998.83778 & 0.11331286 & 0.06667920 & & 0.00432868 \\
		26 & & 1833.50096 & 0.29898848 & 0.06191242 & & 0.00359029 \\
		27 & & 1668.54327 & 0.50434975 & 0.04245542 & & 0.00212884 \\
		28 & & 1593.85828 & 0.61394512 & 0.02313033 & & 0.00051620 \\
		29 & & 1510.37809 & 0.73180548 & 0.00545920 & & 0.00007704 \\
		30 & & 1503.25052 & 0.74518445 & 0.00229869 & & 0.00000069 \\
		31 & & 1500.00008 & 0.74999490 & 0.00150227 & & 2.0564$\times10^{-10}$ \\
		32 & & 1500 & 0.74999999 & 0.0015 & & 5.2284$\times10^{-19}$ \\
		33 & & 1500 & 0.75 & 0.0015 & & 1.3276$\times10^{-26}$ \\
 		\bottomrule
	\end{tabular}
	\caption{\small
		Iterations of inequality-constrained Newton's method performed on the one-layer case using Hessian modifications from starting values $\boldsymbol{\beta}^{(1)} = [2400,1,0.2]^t$\,, converging to true values $[\boldsymbol{\beta}]_1 = [1500,0.75,0.0015]^t$\,.}
	\label{tab:PosBar_beta_2400_1_0.2}
\end{table}

\end{appendix}
\end{document}